\documentclass[usegraphicx,usenatbib,usedcolumn,useAMS]{mn2e}
\newcommand{\per}[1][1]{\scriptsize$^{-#1}$\normalsize}
\newcommand{\kms}{km~s\per}
\newcommand{\unisim}{\sim\!}
\newcommand{\asymerr}[2]{\tiny\makebox[0pt][l]{\raisebox{-0.75ex}{$-#1$}}%
	\raisebox{+1.25ex}{\makebox[0pt][l]{$+$}\phantom{$-$}$#2$}}
\newcommand{\slantfrac}[2]{#1\!\left/#2\right.}
\newlength{\photinfocolwidth}
\newcommand{\bandno}{--}
\newcommand{\bandyes}{$\bullet$}
\newcommand{\bandrf}{$\star$}
\newcommand{\incground}{$\circ$}
\newcommand{\inchst}{$\bullet$}
\newcommand{\rdground}{$\circ$}
\newcommand{\rdhst}{$\bullet$}
\newcommand{\memyes}{C}
\newcommand{\memno}{F}
\newcommand{\noweight}{--}
\newcommand{\clg}[4][ClG]{\mbox{#1 $#2\!#3\!#4$}}
\newcommand{\vrot}{V_{\rmn{rot}}}
\newcommand{\rdspec}{r_{\rmn{d,spec}}}
\newcommand{\rdphot}{r_{\rmn{d,phot}}}
\newcommand{\zcl}{z_{\rmn{cl}}}
\newcommand{\dtf}{\Delta {M_B}^{\!\rmn{TF}}}
\newcommand{\snr}{$\slantfrac{S}{N}$}
\newcommand{\rextent}{r_{\rmn{extent}}}
\newcommand{\papertwo}{Paper~2}
\newcommand{\elfitpy}{\textsc{elfit2py}}
\newcommand{\elfitd}{\textsc{elfit2d}}
\newcommand{\gimd}{\textsc{gim2d}}
\newcommand{\sextractor}{\textsc{SExtractor}}
\newcommand{\pyraf}{\textsc{pyraf}}
\newcommand{\iraf}{\textsc{iraf}}

\title[The TFR of distant cluster galaxies]%
      {The Tully--Fisher relation of distant cluster galaxies%
       \thanks{Based on observations made with ESO Telescopes at
               Paranal Observatory under programme IDs 066.A-0376
               and 069.A-0312.}%
       \thanks{Based on observations made with the NASA/ESA Hubble
               Space Telescope, obtained from the data archive at the Space 
               Telescope Institute. STScI is operated by the association of 
               Universities for Research in Astronomy, Inc. under the NASA 
               contract NAS 5-26555.}}
\author[S. P. Bamford et al.]{%
  S. P. Bamford$^{1}$\thanks{E-mail: ppxspb@nottingham.ac.uk},
  B. Milvang-Jensen$^{2}$, A. Arag\'on-Salamanca$^{1}$
  and L. Simard$^{3}$\\
  $^{1}$School of Physics and Astronomy, University of Nottingham, 
	University Park, Nottingham, NG7 2RD, UK\\
  $^{2}$Max-Planck-Institut f\"ur extraterrestrische Physik, 
	Giessenbachstra\ss e, 85748 Garching, Germany\\
  $^{3}$Herzberg Institute of Astrophysics, 
  	National Research Council of Canada, 
  	Victoria, BC V9E 2E7, Canada
}

\begin{document}
  
\date{Accepted ???. Received ???; in original form ???}

\pagerange{\pageref{firstpage}--\pageref{lastpage}} \pubyear{2005}

\maketitle

\label{firstpage}

\begin{abstract}
We have measured maximum rotation velocities ($\vrot$) for a sample
of $111$ emission-line galaxies with $0.1 \la z \la 1$, observed in
the fields of 6 clusters.  From these data we construct `matched'
samples of $58$ field and $22$ cluster galaxies, covering similar
ranges in redshift ($0.25 \leq z \le 1.0$) and luminosity ($M_B \leq
-19.5$ mag), and selected in a homogeneous manner.  We find the
distributions of $M_B$, $\vrot$, and scalelength, to be very similar
for the two samples.  However, using the Tully--Fisher relation (TFR)
we find that cluster galaxies are systematically offset with respect
to the field sample by $-0.7\pm0.2$ mag. This offset is significant at
$3\sigma$ and persists when we account for an evolution of the field
TFR with redshift.  Extensive tests are performed to investigate
potential differences between the measured emission lines and derived
rotation curves of the cluster and field samples.  However, no such
differences which could affect the derived $\vrot$ values and account
for the offset are found.

The most likely explanation for the TFR offset is that giant spiral
galaxies in distant clusters are on average brighter, for a given rotation
velocity, than those in the field.  This could be due to enhanced
star-formation caused by an initial interaction with the intra-cluster
medium.
As our selection favours galaxies with strong emission lines, this
effect may not apply to the entire cluster spiral population, but does
imply that strongly star-forming spirals in clusters are more luminous
than those in the field, and possibly have higher star-formation rates.
However, the possibility that this TFR offset is a
mass-related effect, e.g. due to the stripping of galaxy dark matter
haloes, is not excluded by our data.

\end{abstract}


\section{Introduction}

The effects of falling into a cluster upon an individual galaxy are
important for a complete understanding of galaxy evolution.  Despite
only a small minority of galaxies being located in rich clusters, even
at zero redshift, such environments are naturally very interesting to
study as extrema.  In particular they are sites of simultaneously both
unusually fast and slow galaxy evolution, and hence contain unique
galaxy populations.  Because of this, they have the potential to
provide much insight into a variety of astrophysical processes, not
only specific to clusters, but also occurring in the general galaxy
population.

A substantial fraction ($\unisim 80$\%) of bright galaxies ($M_b <
-19.5$) in local clusters have no significant current star-formation,
as judged from H$\alpha$ emission \citep{Betal04}.  In addition,
clusters predominantly contain galaxies with elliptical and S0
morphology (again $\unisim 80$\%; \citealt{D80}).  Both of these
observations are in contrast to the local field, for which the same
studies find $\la 40$\% of galaxies to be non-starforming and an
early-type fraction of $\unisim 20$\%.

While some galaxies may have formed in dense regions, it is generally
considered very difficult to create discs under such conditions, as
the cluster environment removes the supply of cold gas from which a
disc might form \citep{GG72}.  In addition the structure formation
scenario of $\Lambda$CDM implies that many galaxies have undergone the
transition from field to cluster environment since $z\la1$
\citep{dLKSWLSTY04}.  At least some of these galaxies must have been
transformed following their entry into the cluster environment, in
order to account for the disparity between the cluster and field
galaxy populations seen today.

The fraction of elliptical galaxies in clusters is observed to be
fairly constant out to $z \sim 1$.  
However, it is well established that the general properties of the
disc galaxy population in distant clusters are different to those
locally, and that a smooth change in these properties can be traced
with redshift, albeit with substantial scatter.  There is a larger
fraction of blue galaxies at high redshift \citep{BO78}, found to be
star-forming \citep{DG82, DG92} and typically with spiral morphology
\citep{CBSES98}. This is in contrast with the quiescent S0 galaxies
which form a significant fraction of the cluster population at low
redshift, and dominate the cores of rich clusters \citep{DOCSEBBPS97}.

A possible implication of all this evidence is that star-forming
spirals are transformed into passive lenticulars by the cluster
environment, and that this is the dominant path for forming such
galaxies, at least in clusters.  While there is evidence at low
redshifts that group environments may be the most important regions
for decreasing the \emph{global} star formation rate of the universe
\citep{Betal04}, for massive galaxies and earlier epochs clusters seem
to be more effective.  Additional evidence for the reality of the
transformation of spirals into S0s is provided by the existence in
clusters of two unusual galaxy types.  The first is passive spirals,
with spiral morphology but no sign of current star-formation.  These
are found in the outskirts of low-redshift clusters, but not generally
in the field, and suggest that some interaction with the cluster
environment has recently curtailed their star-formation
\citep{Getal03}.  The second type are disc galaxies with spectra
indicative of a recent, sudden truncation of their star-formation.
Such galaxies have an E+A spectral type (also known as k+a) with
features of both an old ($>$ several Gyr) and intermediate age ($\la
1$ Gyr) stellar population, but with no sign of on-going
star-formation \citep{DG83, DSPBCEO99}.  Furthermore, many such
spectra indicate a star-burst occurred shortly prior to the end of
star-formation \citep{PSDCBBEO99}. E+A galaxies are found over a wide
redshift range, but in local clusters most are dwarfs
\citep{PBKYCMOK04} and in the field they have almost entirely
elliptical or irregular morphologies \citep{YZZLM04,TFIvDKM04}.  The
larger, disky E+As which may form the link between spirals and S0s are
preferentially found in clusters at intermediate redshifts, where the
relative fraction of spirals and S0s is seen to change most rapidly.

Several mechanisms have been proposed that could transform spirals to
S0s in cluster environments.  The favoured options are
ram-pressure stripping \citep{GG72} and unequal-mass mergers
\citep{B98,MH94}.

In the ram-pressure stripping scenario, the pressure due to the
galaxy's passage through the intra-cluster medium (ICM) removes gas
that would have fuelled future star-formation.  Depending upon the
model one assumes, the gas could be removed from the disc itself,
causing a fast truncation of star-formation \citep{AMB99,QMB00}, or
the gas could merely be removed from the galaxy halo \citep{BCS02}.
Normally the disc gas consumed in star-formation is replenished by
infall from the reservoir of halo gas. This latter alternative thus
leads to a gradual decline in star-formation rate (SFR) as the
quantity of available disc gas diminishes.  Prior to the cessation of
star-formation, the increased pressure in the disc gas may actually
trigger an initial burst of star-formation, through compression of the
galactic molecular clouds \citep{BC03}.  This in turn would cause an
increase in the rate of disc gas consumption, and hence reduce the
time taken for star-formation to cease.  The duration of any
star-burst of this form is therefore self-limiting, and necessarily
short, with the strongest bursts being the shortest-lived.

Galaxy mergers may cause an eventual truncation of star-formation by
first inducing a star-burst.  This enhanced SFR quickly depletes the
supply of gas from which future stars could have formed, and thus
subsequently halts star-formation.  Gas from the outer disc is also
tidally stripped, reducing the amount available for
star-formation. From simulations, \citet{B98} find that mergers with a
mass ratio of $\unisim3\!:\!1$ often result in S0 morphologies.  Minor
mergers (mass ratio $\ga10\!:\!1$) have a smaller effect on the larger
galaxy, the disc is dynamically heated and therefore becomes thicker,
but repeated minor mergers may lead to an S0 appearance.  Mergers
between galaxies of nearly-equal mass, on the other hand, while also
inducing a star-burst and consequent end of star-formation, generally
destroy any disc component, resulting in an elliptical morphology.

Both of these mechanisms may well occur, and result in galaxies with
roughly S0 morphology and typically corresponding spectral properties.
However, we would like to know which has the dominant role, and
examine any differences in the form of the transformations, including
how E+A galaxies fit into the evolution.  In addition there is likely
to be a dependence of the S0 formation mechanism on environment, which
deserves attention.  For example, the high relative velocities in
clusters make merging less likely than in groups, while ram-pressure
stripping is probably only effective in the dense ICM of large
clusters.

Another potential effect, present in clusters, is galaxy harassment
\citep{MLQS99}.  This is caused by the tidal effects of close
encounters with other, more massive, galaxies.  However, while this
may contribute somewhat to a thickening of discs in clusters, it is
more important for dwarf galaxies than the giant discs we are
considering here.  A tidal effect likely to be more significant for
the evolution of massive galaxies is the tidal field due to the
cluster potential itself.  While in the smooth, static case this is
judged to only be important close to the cluster core \citep{HB96},
the existence of substructure, and in particular cluster-group and
cluster-cluster mergers, may result in a time-varying tidal field with
more significant effects \citep{B99,G03a,G03b}.  \citet{OLKWM05} claim
that this is the most likely explanation for their observations of
enhanced star-formation in Abell 2125 (at $z=0.25$), which appears to
be undergoing a cluster-cluster merger.

A potential key difference between the transformation by ram-pressure
stripping and through mergers or tidal effects is that the former is
likely to enhance star-formation across the disc \citep{BC03}, while
any star-burst caused by merging or tides is probably centrally
concentrated, due to disc gas being driven inward by an induced
central bar \citep{MH94}.  These differences may be distinguishable,
once a luminosity enhancement has been established for a galaxy, by
examining colour gradients or more detailed properties of the
stellar-populations as a function of radius.  For example, if the
galaxy centre is bluer than the disk, this implies centrally enhanced
star-formation, and therefore possibly that a tidal interaction is
responsible.  However, such interpretations will require careful
comparison with simulations of galaxies' internal responses to the
various mechanisms.  We do not attempt to examine colour gradients in
our present sample, due to the uncertainties that would be caused by
the heterogeneous nature of our imaging.

In complement to the examination of stellar-population gradients,
differences in the time-scales of the star-burst and subsequent SFR
decline may also help to distinguish between the proposed mechanisms.

To summarize, much evidence has been accrued for the transformation of
spirals to S0s by the cluster environment, and a number of plausible
mechanisms have been proposed, but there is still little known about
its detailed nature and few constraints on which mechanism is actually
responsible.  We are conducting a study to address these issues using
a variety of techniques.  In this paper we examine the first stage of
this phenomenon, the early effect on spiral galaxies falling into a
cluster.  By comparing the Tully--Fisher relation (TFR) for field and
cluster galaxies, we aim to evaluate the cluster's effect upon the
mass-to-light ratio of a galaxy during the period for which it retains
spiral morphology and an appreciable star-formation rate.  Assuming
star-formation is eventually suppressed in cluster galaxies, such
galaxies are thus presumably recently arrived from the field.  We can
therefore investigate the existence and prevalence of luminosity (and
hence perhaps star-formation rate) enhancement in the early stages of
the spiral to S0 transformation.

Our initial work, a pilot study of one cluster, MS1054 at $z=0.83$
(\citealt{MJetal03}, described in more detail in \citealt{MJthesis}
and Milvang-Jensen \& Arag{\' o}n-Salamanca in preparation) found
evidence for a $B$-band brightening of the cluster population with
respect to the field of $\unisim 1$ mag.  Following the success of
this work, we have embarked upon a larger study of nine additional
clusters.  For this programme we have employed optical multi-slit
spectrographs on two telescopes; six of our clusters (including
MS1054) were observed using
FORS2\footnote{\label{fn:fors}http://www.eso.org/instruments/fors} on
the VLT \citep{FORS}, and four with
FOCAS\footnote{http://www.naoj.org/Observing/Instruments/FOCAS} on
Subaru \citep{FOCAS}.  These two data sets have been separately
reduced and analysed, in order to provide semi-independent tests and
allow evaluation of the robustness of any results to different
reduction methods.  
This paper considers the clusters observed using VLT/FORS2, including
the data for MS1054 studied previously.  While the acquisition and
reduction of the MS1054 data differs slightly from the other VLT/FORS2
clusters, the details do not warrant repetition here.  Basic details
of the clusters discussed in this paper are given in Table
\ref{tab:clusters}.  The Subaru/FOCAS data will be the subject of a
subsequent paper (Nakamura et al. in preparation).

\begin{table*}
\begin{minipage}{0.69\textwidth}
\caption{\label{tab:clusters} The positions, redshifts, velocity
dispersions and adopted redshift ranges of the clusters observed for
this study using FORS2 at the VLT.  The alternative names are those
preferred by Simbad (http://simbad.u-strasbg.fr) at CDS, following IAU
recommendations.  For each field, galaxies with $\zcl - \Delta
\zcl \le z \le \zcl + \Delta \zcl$ are considered to be
cluster members.}
\begin{tabular}{llccccc}
\hline
Cluster & Full alt. name & R.A. [J2000] & Dec. [J2000] &
	$\sigma$ [\kms] & $\zcl$ & $\Delta \zcl$\\
\hline
MS0440$^{\,\rmn{a}}$  & \clg{0440}{+}{02} & 04 43 09.5 &
    $+02$ 10 30 & 838  & 0.197 & 0.010\\
AC114$^{\,\rmn{a}}$   & ACO S 1077        & 22 58 47.1 &
    $-34$ 47 60 & 1388 & 0.315 & 0.018\\
A370$^{\,\rmn{a}}$    & ACO 370           & 02 39 51.6 &
    $-01$ 34 12 & 859  & 0.374 & 0.012\\
CL0054$^{\,\rmn{a}}$  & \clg{0054}{-}{27} & 00 56 56.0 &
    $-27$ 40 32 & 742  & 0.560 & 0.012\\
MS2053$^{\,\rmn{b}}$  & \clg{2053}{-}{04} & 20 53 44.6 &
    $-04$ 49 16 & 817  & 0.583 & 0.013\\
\hline
MS1054$^{\,\rmn{a}}$  & \clg{1054}{-}{03} & 10 56 57.3 &
    $-03$ 37 44 & 1178 & 0.830 & 0.022\\
\hline
\end{tabular}
\\
$^{\rmn{a}}$ position, $z$ and $\sigma$ from \citet{GM01}.\\
$^{\rmn{b}}$ position and $z$ from \citet{SMGMSWFH91},
$\sigma$ from \citet{HFKvD02}.
\end{minipage}
\end{table*}

In addition to cluster galaxies, we observe a large number of field
galaxies for comparison, with redshifts in the range $0 \la z \la 1$.
These form a useful sample for evaluating galaxy evolution purely in
the field, and are used for this purpose in
\citet*[hereafter \papertwo]{BamfordField}.

In \S2 we describe our target selection procedure and summarize the
properties of the objects observed. Section 3 gives details of the
photometric and spectroscopic data employed in this study, and
describes the methods used to derive galaxy parameters from them. The
general properties of the cluster and field samples are contrasted,
and matched samples constructed in \S4, and the TFR is considered in
\S5. In \S6 we discuss our findings and compare with other recent
works, and present our conclusions in \S7.  Throughout we assume the
concordance cosmology, with $\Omega_{\Lambda} = 0.7$, $\Omega_{m} =0.3$
and $H_0 = 70$ \kms~Mpc\per{} \citep{WMAP}.  All magnitudes are in
the Vega zero-point system.

\vspace{-1.2\baselineskip}
\section{Target selection}

The clusters in our sample were simply selected to be rich clusters
covering a wide redshift range and with available HST imaging, and
therefore do not form a particularly homogeneous sample.  However,
this is not regarded as a problem for our purposes, as we are
primarily seeking to establish the reality of a difference between
cluster and field galaxies, and gain a first insight into the nature
of any disparity.  We leave a detailed examination with respect to
cluster properties, redshift, etc.\ for larger, more homogeneous
studies such as the ESO Distant Cluster Survey (EDisCS;
\citealt{EDISCS_Messenger,EDISCS}).

The galaxies observed within each field were selected by assigning
priorities based upon the likelihood of being able to measure a
rotation curve, making use of any previously known spectral properties
(MS0440: \citealt{GSLFFLH98},
AC114: \citealt{CS87,CBSES98},
A370: \citealt{DSPBCEO99,SDCEOBS97},
CL0054: \citealt{DSPBCEO99,SDCEOBS97}; P-A Duc private comm.,
MS1054: \citealt{vDFFIK00}).
Initial catalogues were constructed from the $R$-band preimaging, with
each galaxy being given priority 5 (lowest).  The priority level of
each galaxy was then decreased by one point for each of the following:
disky morphology, favourable inclination, known emission line
spectrum, and available HST data.  A priority level of 5 was assigned
to all galaxies close to face-on ($i < 30\degr$).  The galaxies were
thus divided into five priority categories from 1 (highest) to 5
(lowest). The aim was to select field and cluster galaxies in as
similar a way as feasible, that is while still observing a useful
number of galaxies actually in the cluster.  To increase the
likelihood of observing cluster galaxies priority was also increased
by one point if the galaxy was known to be at the cluster redshift and
did not already have the highest priority level.

Our priority ranking method preferentially selects bright,
star-forming disc galaxies, and therefore we are not probing the
average spiral population in clusters.  However, by selecting field
galaxies in the same manner we can perform a fair comparison between
the bright, star-forming population in clusters and the corresponding
population in the field.  We can therefore investigate whether there
is any evidence for a brightening or fading of this population in
clusters.

For each mask, slits were added in order of priority, and within each
priority level in order from brightest to faintest $R$-band magnitude.
The only reason for a particular galaxy not being included is a
geometric constraint caused by a galaxy of higher priority level, or a
brighter galaxy in the same priority level. Often the vast majority of
the mask was filled with slits on galaxies in priority levels 1 and 2,
with occasional recourse to lower priority objects in order to fill
otherwise unoccupied gaps.  The effective magnitude limit in each
priority level varies, and is generally limited by either the
availability of spectroscopic data or slit positioning constraints.

\begin{figure}%
\centering%
\includegraphics[height=0.45\textwidth,angle=270]%
		{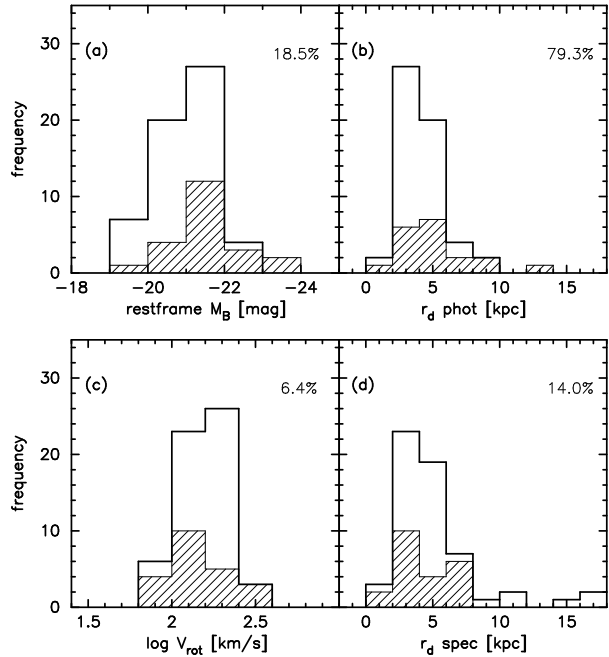}%
\caption{\label{fig:param_distrib}%
	The distributions of
	(a) absolute rest-frame $B$-band magnitude ($M_B$), 
	(b) photometric disc scalelength ($\rdphot$),
	(c) rotation velocity ($\vrot$), and
	(d) spectroscopic emission scalelength ($\rdspec$)
	for the galaxies in our final `matched' TFR sample.
	The empty histogram corresponds to field galaxies,
	while the hatched histogram corresponds to 
	cluster members.
	The percentage in the top right corner of each panel
	indicates the confidence that the field and cluster 
	samples are drawn from the same distribution, as given by
	a K-S test.
	}
\end{figure}

As the multi-object spectroscopy limits the number and minimum
separation of targets, the observed galaxies are rather sparsely
sampled.  As shown in Fig.~\ref{fig:param_distrib}, the preference for
cluster galaxies does therefore not significantly extend nor bias the
parameter space inhabited by the cluster galaxies with respect to that
of the field galaxies.  It merely means that cluster galaxies are
slightly over-represented compared with a purely magnitude limited
sample.  We can therefore internally evaluate the difference between
cluster and field galaxies over a range of redshifts, using galaxies
that have been selected, observed and analysed in an essentially
identical manner. We have no need to resort to comparisons with other
studies, and hence avoid the systematic differences this could
potentially involve.

\begin{figure}%
\centering%
\includegraphics[height=0.45\textwidth,angle=270]%
		{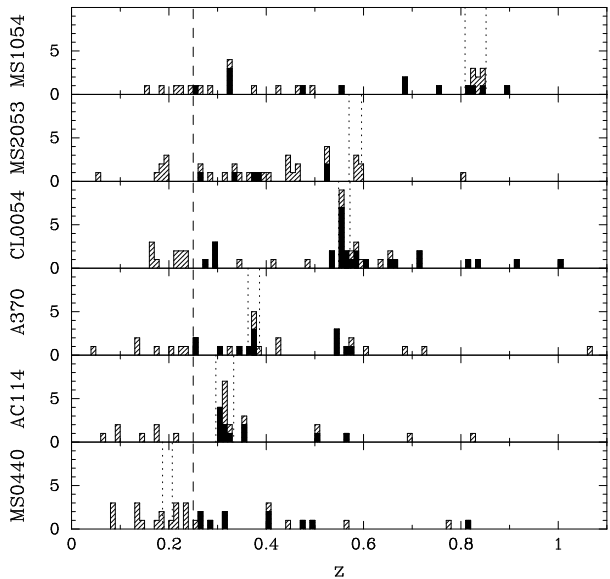}%
\caption{\label{fig:zdist}%
	The redshift distribution of our target galaxies.
	The hatched histogram gives the distribution of all our
	observed galaxies with identifiable emission lines.
	The filled area shows only those in our final `matched'
	TFR sample. Vertical dotted lines indicate the adopted
	cluster limits, and the dashed line shows the low redshift
	cut for our `matched' sample.}
\end{figure}

The redshift distributions of our sample galaxies are shown in
Fig.~\ref{fig:zdist}.  Clearly the number of galaxies selected which
actually lie in the targeted cluster varies considerably between the
observed fields.  This is primarily a consequence of variation in the
spiral population of the clusters, and differing availability of
\emph{a priori} redshifts during the target selection.  Note that the
shorter exposure time required for MS0440 meant we could use three
masks, compared with two for the other clusters.  The low numbers of
selected cluster galaxies, although unfortunate, does go some way to
demonstrate the extent to which we have endeavoured to keep our sample
unbiased.

In order to best observe the galaxy kinematics, the slits were
individually tilted to align with the major axes of the target
galaxies.  Tilting the slits reduces the effective spatial resolution,
and so multiple masks, with different position angles on the sky, are
required to accommodate all galaxy position angles. We generally used
two, orthogonally aligned, masks for each field, and
thus a nominal limiting slit tilt of $45\degr$.  From previous work we
have found that useful spectra can be obtained using slits tilted up
to this limit.  However, on occasion the $45\degr$ limit was exceeded,
principally in order to observe the same object in both masks for
comparison.  For the three MS0440 masks the same tilt limit was
applied to maximise the number of high priority targets which could be
fit in the masks.
In the completed designs of the 11 masks (not including MS1054), $283$
slits were assigned to targets, including $34$ stars (roughly three
per mask) for the purposes of alignment and measuring the seeing.

\vspace{-1.2\baselineskip}
\section{Data}

\subsection{Photometry} \label{sec:phot}

The photometry used in this study is primarily from the HST archive
and our own FORS2 $R$-band imaging.  These are supplemented with
additional reduced and zero-point calibrated ground-based data, kindly
provided by Dr.~Ian Smail, in order to provide colour information.
This additional colour information is advantageous for constraining
the galaxy SED and hence improving the $k$-correction.  The
photometric zero-points for our $R$-band imaging were established by
matching the magnitudes of point-sources with those measured on the
overlapping $\unisim R$-band ($F675W$ or $F702W$) HST images.  In one
case (CL0054) no $\unisim R$-band HST data were available, and an
interpolation between $F555W$ and $F814W$ magnitudes was calibrated
using synthetic SEDs and used instead.  This was also checked using
$V$ and $I$-band ground-based data, which gives a consistent
zero-point.  The mean zero-point error on the $R$-band magnitudes is
$0.08$ mag, adequate for our purposes.  The zero-point errors are
included in the overall magnitude errors.
Table \ref{tab:phot_info} gives the bands in which magnitudes were
measured for each galaxy in our full TFR sample (as defined later).

The galaxy magnitudes were measured using \sextractor{}
\citep{sextractor}.  The AUTO (Kron-style) aperture was used to
measure magnitudes on the original images, while colours were
determined from 3 arcsec diameter aperture magnitudes measured on
images which had been degraded to match the worst seeing for each
field.  Magnitudes and colours were corrected for Galactic extinction
using the maps and conversions of \citet*{SFD98}.

The conversion from apparent magnitudes to absolute rest-frame
$B$-band was achieved by the following procedure.  First all colour
information was used to find the best fitting SED from a grid of 26,
spanning types E/S0 to Sdm. These were formed by interpolating the
SEDs of \citet{AECC93} and redshifting appropriately.  A confidence
interval on the SED was also determined by examining the $\chi^2$ of
the colour fits. In cases where no colour information was available an
average value and confidence interval were adopted for the SED,
determined from those galaxies with available colours.  The magnitude
in the observed band closest to rest-frame $B$ was then adjusted by a
colour- and $k$-correction calculated from the best-fitting
SED. 
The observed band used as the basis for the conversion is indicated
in Table \ref{tab:phot_info}.
Errors were assigned to this correction corresponding to the SEDs
bounding the confidence interval determined above.  Finally this
magnitude was adjusted by the distance modulus of the galaxy assuming
the concordance cosmology.

Because of the varying imaging available for each galaxy, we need to
address concerns that $M_B$ for our cluster galaxies may be
systematically biased with respect to the field galaxies; for example,
due to different colours being available to determine the SED.  Our
additional imaging tends to be centred on the cluster and,
particularly for the HST imaging, often has a smaller field-of-view
than the $R$-band images from which we selected the targets.  Cluster
galaxies will tend to be located towards the centre of the
field-of-view (although this may not be so true for the star-forming
spirals in our sample, particularly given that we are confined to a
clustercentric radius of $\la1$ Mpc).  We might thus expect cluster
galaxies to have imaging available in more bands.  However, on average
we have a magnitude measured in $2.2$ and $2.1$ bands for cluster and
field galaxies respectively, so there is no evidence for a difference
in the number of colours available.  Another concern may be that the
apparent magnitude used as a basis for $M_B$ is measured on HST images
more often for cluster galaxies than for field galaxies.  The opposite
is actually the case, ($14\pm8$)\% of cluster galaxies have $M_B$
based on HST imaging, compared with ($29\pm6$)\% of field galaxies.
However, this is not especially significant, given the Poisson errors.
These tests imply that any significant difference measured between the
cluster and field samples cannot be attributed to the heterogeneity of
the imaging.

The magnitudes were additionally corrected for internal extinction
(including face-on extinction of $0.27$ mag), following the
prescription of \citet{TF85}, to give the corrected absolute
rest-frame $B$-band magnitudes, $M_B$, used in the following analysis.

Note that both the cosmology and internal extinction correction
prescription were chosen to allow straightforward comparison with
other recent studies.

Inclinations and photometric scalelengths for the disc components of
the observed galaxies were measured by fitting the galaxies with
bulge+disc models using the \gimd{} software developed by
\citet{GIM2D}.  This was done in all available imaging as a
consistency check, and all the fits were inspected to determine the
(infrequent) occasions where \gimd{} failed to correctly fit the
data.  

The inclinations, $i$, used in this study are generally those measured
in the HST images by \gimd.  For galaxies without HST imaging, or with
no acceptable HST fits, the best available inclination, from
ground-based \gimd{} fits or \sextractor{} axial ratios was used.
These supplemental inclinations were examined for biases, using a set
of galaxies with both HST and ground based measurements, and corrected
to match the HST \gimd{} values.  Errors on the inclinations were
estimated from the scatter in values derived from different images and
methods.

Photometric disc scalelengths, $\rdphot$, were derived from \gimd{}
fits, preferentially on HST imaging, but also on ground-based images
if HST fits were not available.  Only bands near to $R$ were used to
reduce any contamination due to a potential variation in scalelength
with observed wavelength \citep[e.g.][]{dJ96}.  A bias with respect to
rest-frame wavelength may remain, but does not affect this study
significantly.  Scalelengths are given in kpc, calculated assuming the
concordance cosmology.
The final two columns in Table \ref{tab:phot_info} indicate whether
the inclinations and photometric disc scalelengths were measured on HST
or ground-based imaging.

Many of our photometric disc scalelengths are necessarily measured on
ground-based images, often with seeing $\ga 1$ arcsec FWHM.  At
$z=0.5$ the angular scale is $6.1$ kpc arcsec\per{} in our adopted
cosmology. For our sample $\rdphot$ is typically $\unisim 4$ kpc, and
thus the FWHM of the disk surface brightness profile is $\unisim 4
\times 2\ln{2} \simeq 6$ kpc.  This is a comparable scale to the
seeing, making the measurement of $\rdphot$ difficult.  For a
single-component surface brightness we would expect the \gimd{}
measurement of $\rdphot$ to be reliable, although sensitive to the
precision of the seeing determination.  However, with a more
complicated, realistic surface brightness profile, being fit by
two-component model, the limited resolution makes measurement of the
disk scalelength potentially unreliable.  This problem becomes
significantly worse for higher redshifts and smaller galaxies.  We
therefore prefer not to base any inferences upon our $\rdphot$
measurements and do not consider them further in this paper, except in
rough comparisons of the cluster and field parameter distributions,
and as an aid to the emission line quality control procedure described
in the following section.

\vspace{-0.5\baselineskip}
\subsection{Spectroscopy} \label{sec:spec}

The multi-slit spectroscopy for this study was observed using the
MXU\footnotemark[1] mode of FORS2 on the VLT.  In this
mode slits are cut into a mask which is then placed in the light path.
This has significant advantages over the movable slits of
MOS\footnotemark[1] mode. Variable slit lengths and tilt
angles give increased flexibility for the mask design, increasing the
number of objects observable in a single exposure and allowing
consistent alignment of the slits with the galaxy major axes.  Our
observations are summarized in Table \ref{tab:obs}.  The seeing, as
measured from stellar spectra in the masks, was typically $\unisim 1$
arcsec, and always less than $1.2$ arcsec.  The setup was similar to
that used for the earlier MS1054 observations (an additional 2 masks),
the only changes being a larger CCD detector and a different grism
(600RI) with a substantially higher throughput.  These differences
give a wider wavelength coverage, although with a slightly lower
spectral resolution, meaning more emission lines were observed for
each galaxy in the present study.

\begin{table}%
\caption{\label{tab:obs}%
	Summary of our FORS2 spectroscopic observations.}%
\centering%
\begin{tabular}{@{\hspace{0.5cm}}lcc@{\hspace{0.5cm}}}
\hline
Cluster-field & No. of masks & Exp.~time (mins/mask)\\
\hline
MS0440  &  3  &  57, 45, 30  \\
AC114   &  2  &  75          \\
A370    &  2  &  90          \\
CL0054  &  2  &  150         \\
MS2053  &  2  &  150         \\
\hline
MS1054  &  2  &  210         \\
\hline
\end{tabular}
\end{table}

Our spectroscopic data were reduced in the usual manner
(\citealt{MJthesis}; Bamford thesis in preparation), using
\pyraf%
\footnote{PyRAF is a product of the Space Telescope Science Institute,
          which is operated by AURA for NASA.}
to script standard \iraf{} tasks and our own routines, producing
straightened, flat-fielded, wavelength-calibrated and sky-subtracted
2d-spectra.  The main emission lines observed were [OII]$\lambda
3727$, H$\beta$ and [OIII]$\lambda\lambda4959,5007$, with H$\alpha$
but no [OII]$\lambda 3727$ for nearby galaxies.  Emission lines were
identified by eye and small regions of the spectra containing each
line extracted.  The continuum emission was subtracted from each of
these small images, which were then cropped further to produce
`postage stamps' of each line.
In the $283$ slits (still not including the MS1054 observations),
$303$ separate spectra were identified.  Of these $177$ are
identifiable as galaxies with emission lines.
Note that from the $20$ serendipitously observed galaxies, only $2$
field galaxies that happened to be well aligned with the slit, and met
all the other criteria, have been included in the TFR study.

In order to measure the rotation velocity ($\vrot$) and emission scale
lengths ($\rdspec$) we fit each emission line independently using a
synthetic rotation curve method based on \elfitd{} by
\citet{SP98,SP99}, and dubbed \elfitpy\footnote{The main differences
between the method of \citet{SP99} (\elfitd) and that used here
(\elfitpy) are a $4\times$ spectral oversampling to reduce the
velocity `quantisation' found by \citet{MJthesis}, the use of an error
image rather than a constant noise level, improving performance in the
vicinity of skyline residuals, and the addition of a test to judge
when convergence has been achieved.}.  In this technique model
emission lines are created for particular sets of parameters, and
compared to the data to assess their goodness-of fit.  The model
emission lines are created assuming a form for the intrinsic rotation
curve, an exponential surface-brightness profile, and given the galaxy
inclination, seeing and instrumental profile.  The intrinsic rotation
curve assumed here is the `universal rotation curve' (URC) of
\citet{PS91}, with a slope weakly parametrized by the absolute
$B$-band magnitude, $M_B$.  Adopting a flat rotation curve leads to
values of $\vrot$ $\unisim 10$ \kms{} lower, but does not affect the
conclusions of this study.
As well as $\vrot$ and $\rdspec$, the emission line flux, constant
background level and, in the case of [OII], the doublet line ratio,
are simultaneously fit.

A Metropolis algorithm (\citealt{MRRTT53}, as described by
\citealt{SW94}) is used to search the parameter space to find those
which best fit the data, and to determine confidence intervals on
these parameters.  Images of model lines with the best-fitting
parameters are also produced for comparison with the data.

While there may be some concerns about the Metropolis algorithm
finding a local, rather than global, minimum, inspection of the time
series of accepted points in the Metropolis search shows that the
$\vrot$ and $\rdspec$ parameters converge fairly quickly to their
final values, and are usually stable around these values for the
remainder of the sampling iterations.  This implies fairly deep and
smooth global minima in chi-squared space, with few local minima.  

In contrast, the less well constrained, but also less critical,
parameters of background level and doublet ratio show more frequent
jumps between semi-stable values.  While this reveals the existence of
local minima, it also demonstrates the algorithm's ability to move out
of such regions when they exist.  The final error in the measured
parameter thus includes the uncertainty due to the multiplicity of
chi-squared minima.  The jumps between minima in these subsidiary
parameters are rarely accompanied by any significant shift in the
stable $\vrot$ and $\rdspec$ parameter values.

Two of the emission line galaxies do not have absolute $B$-band
magnitudes from the photometry, and a further four have no lines
suitable for fitting (i.e. the lines were so faint that the mean flux
across the postage stamp was negative due to the noise).  An
additional $17$ galaxies were discarded due to their inclinations
being deemed highly uncertain.  For the remaining $153$ objects, a
mean of $3.3$ suitable lines per galaxy were fit by the procedure
described above.

The principal results of the rotation curve fitting are measurements
of $\vrot$ and $\rdspec$ with estimates of their error for, in
general, several emission lines per galaxy.  (Actually $\vrot \sin{i}$
is measured, which is converted to $\vrot$, using the inclinations
described in \S\ref{sec:phot}, once an average value of $\vrot
\sin{i}$ has been determined for each galaxy.)
In order to produce a single value of $\vrot$ and $\rdspec$ for each
galaxy the values for the individual lines (labelled by $j$ below) are
combined by a weighted mean. 
Upper and lower errors ($+,-$) on these average parameters are
determined as the maximum of (a) a weighted combination of the
individual errors estimated by \elfitpy, and (b) the standard error of
the weighted mean determined from the individual measurements.  For
example, with weights
\begin{equation}
w_j = \frac{2}{({\sigma_{\vrot,j}^+}^{\!\!2}+{\sigma_{\vrot,j}^-}^{\!\!2})} \;,
\end{equation}
the computed upper error on the average $\vrot$ is the square-root of
\begin{eqnarray} \label{eqn:vrot_error}
{\sigma_{\vrot}^+}^{\!\!2} = \!\! & \max\Bigg[ \!\!\! &
     \sum_j{w_j^2\,{\sigma_{\vrot,j}^+}^{\!\!2}}\bigg/\,\biggl({\sum_j{w_j}}\biggr)^2
         \;, \\
 & & \sum_j{w_j\left(V_{\rmn{rot},j} - 
                     \overline{\vrot}\right)^2}\biggl/\biggl({n\sum_j{w_j}}\biggr)
  \Bigg] \nonumber
\end{eqnarray}
where $n$ is the number of measurements contributing to the average.
The first term in the $\max$ function corresponds to case (a) above,
and the second to case (b).  
The lower error $\sigma_{\vrot}^-$ is computed similarly.

For most galaxies the error calculated from those given by \elfitpy{}
is close to that inferred from the standard error of the data, and
hence case (a) applies, or case (b) causes a negligible increase in
the error.  However, for galaxies where there is inconsistency between
values from different lines, and no way of determining which lines
should be preferred, case (a) would underestimate the true
uncertainty.  In these situations case (b) provides a more realistic
estimate of the error.  This test is obviously not possible for
galaxies with only one observed (and accepted by the quality control
procedure) emission line, and therefore will cause formally
inconsistent errors.  However, this is judged to be a minor problem
when compared with the elimination of occasional situations where the
uncertainty would otherwise be seriously underestimated.

A significant fraction of the emission lines identified display
dominant nuclear emission, or asymmetries in intensity, spatial extent
or kinematics.  In severe cases these departures from the assumed
surface brightness profile and intrinsic rotation curve mean that the
best-fitting model is not a true good fit to the data.  A similar
situation can occur for very low signal-to-noise (\snr) lines, where
an artifact of the noise overly influences the fit. More concerning is
the case of very compact lines, where the number of pixels is on the
order of the number of degrees of freedom in the model, and hence an
apparently good fit is obtained despite a potentially substantial
departure from the assumed surface brightness profile.

In order to eliminate such `bad' fits a number of quality tests are
imposed, based on a measure of the median \snr{} (per pixel over the
region where the model line has significant flux), and a robust
reduced-$\chi^2$ goodness-of-fit estimate ($\chi_{\rmn{r}}^2$).  The
cuts on these quantities were set following a detailed simultaneous
inspection of the data, model line and best-fitting parameters.
Firstly, for each line a lower limit in \snr{} is applied, followed by
an upper limit on $\chi_{\rmn{r}}^2$.  As an initial attempt at
excluding sources too compact to fit reliably, lines were also
rejected if the best-fitting scalelength of the emission was
consistent with zero within the $1\sigma$ confidence interval derived
by \elfitpy.

These cuts alone were deemed too inefficient, i.e. cuts rejecting all
obviously `bad' fits resulted in an excessive number of clearly `good'
fits being discarded. Ideally we would prefer an entirely quantitative
method, and therefore a number of additional quantities were
calculated to assist the quality judgement.  The line was fit by a
Gaussian in each spatial row%
\footnote{using the \iraf/\textsc{stsdas} task \textsc{ngaussfit}},
with the errors on each Gaussian fit determined by repeated
simulations with different noise realisations corresponding to the
error image.  It was found that for the [OII]$\lambda 3727$ doublet,
fitting a single Gaussian was more robust than attempting to
simultaneously fit both components.  Through inspection of the
parameters and their errors quantitative criteria were developed for
judging whether the fit position is reliable. The emission line
was thus `traced' and the region determined for which the trace is
reliable.

The distance from the continuum centre to where the line could no
longer be reliably detected above the noise we term the \emph{extent}
($\rextent$).  This quantity is dependent on the properties of the
data, e.g. pixel size and seeing, and is thus not suitable for comparison with
other studies.  However, it is useful for the internal investigation
of differences between various subsets of our own data set.  Note that
\elfitpy{} uses all the pixels simultaneously, and therefore
successfully uses information further out than $\rextent$ when fitting
a model line. Indeed, reasonable fits can be obtained even with lines
for which $\rextent$ is zero.

Additional quantities describing the asymmetry, in terms of extent and
kinematics, and the flatness of the line at maximum extent were also
formulated.  However, a satisfactory set of criteria based on these
quantities could not be found.  Therefore, with the above cuts on \snr
and $\chi_{\rmn{r}}^2$ established all of the model lines were
reviewed by eye, along with the galaxy images, observed emission line,
model parameters and the various quantities just described.  From this
inspection lists were compiled of those lines to be unconditionally
excluded or included in the calculation of $\vrot$. The main occasions
where such action was necessary was to exclude lines which were
clearly due to very central emission, judged in combination with the
ratios $\slantfrac{\rdspec}{\rdphot}$ and
$\slantfrac{\rextent}{\rdphot}$, but where $\chi_{\rmn{r}}^2$ was low
enough to make the adopted cut.

Also unconditionally excluded were lines which made the
$\chi_{\rmn{r}}^2$ and \snr{} cuts, but were obviously incorrect or
clearly inconsistent with other lines available for the galaxy,
particularly when this was for an obvious reason such as low \snr{} or
interference from skyline residuals.  The primary cases for
unconditional inclusion were where slight asymmetries and/or
absorption wings caused a high $\chi_{\rmn{r}}^2$ value, but the fit
was clearly well matched to a high \snr{} line with large
$\slantfrac{\rextent}{\rdspec}$ and $\rdspec \ga \rdphot$.

After the application of these visual exclusions and inclusions, any
galaxy with an average $\vrot$ consistent with zero rotation, within
the errors given by Eqn. \ref{eqn:vrot_error}, was discarded from the
sample.  While this is not ideal, it is very useful to remove galaxies
for which the emitting region is probably not rotationally supported.
The above line selections were then re-applied with the additional
constraint that individual lines were also rejected if their
best-fitting $\vrot$ was consistent with zero within the $1\sigma$
confidence interval derived by \elfitpy.

In five cases there are two spectra corresponding to the same galaxy,
both intentionally, for comparison purposes, and coincidentally.  In
one case the second observation is with a slit at a $\unisim30\degr$
angle to the major axis of the galaxy and thus of much lower quality.
This observation was therefore discarded.  The remaining four galaxies
have reasonably consistent measured parameters from their duplicated
spectra and thus weighted averages of the fit parameters are adopted.

After the rigorous line quality-control procedure, $93$ galaxies
remain.  These comprise the TFR sample of the five cluster fields
observed for this study.  In order to consistently combine the MS1054
data with this sample, the emission line postage stamps for the MS1054
galaxies have been re-fit using \elfitpy{} and the same line quality
criteria applied.  Of the $31$ galaxies with emission in the MS1054
observations, $18$ remain after applying the line quality criteria.
These are added to the sample described above, giving a total of $111$
galaxies, with a mean of $2.3$ lines contributing to the measurements
for each galaxy.

\vspace{-0.5\baselineskip}
\subsection{Data table}

The data described above are presented in Table \ref{tab:data}.  The
first four columns give the ID assigned in this study, the R.A.\ and
Dec.\ in J2000 coordinates and the redshift ($z$).  Column 5 (labelled
`Mem.') indicates whether each galaxy is a cluster member according to
the criteria used in this paper, `F' and `C' indicate field and
cluster respectively.  The galaxy inclination is given in column 6
($i$), with $90\degr$ corresponding to edge-on, and column 7
($A^{\rmn{i}}$) lists the internal extinction corrections applied
(including 0.27 mag of face-on extinction), with errors corresponding
to the uncertainty in the measured inclination.  Column 8 gives the
absolute rest-frame $B$-band magnitude ($M_B$) in our assumed
cosmology ($\Omega_{\Lambda} = 0.7$, $\Omega_{m} =0.3$, $H_0 = 70$
\kms~Mpc\per{}), corrected for Galactic and internal extinction, with
errors including the contributions from the initial measurement and
all subsequent corrections.  Column 9 specifies the rotation velocity
($\vrot$) for each galaxy, derived from \elfitpy{}, averaged over all
lines which pass the quality control criteria.  The errors on this
quantity include those determined via Eqn. \ref{eqn:vrot_error},
combined with the error on the correction from $\vrot \sin{i}$ to
$\vrot$. Finally, column 10 gives the weight ($w_\rmn{TF}$) assigned
to each galaxy in the TFR fits of Eqn. \ref{eqn:tf_fits} in \S5.  Only
galaxies in the matched samples, defined below in \S\ref{sec:matched},
are included in these TFR fits, and therefore only these galaxies have
$w_\rmn{TF}$ values.

\begin{table*}%
\begin{minipage}{0.88\textwidth}%
\caption{\label{tab:data}%
	The data for our full field and cluster TFR samples.
	The columns are:
	(1)~ID assigned in this study,
	(2)~R.A.\ and (3)~Dec.,
	(4)~redshift, (5)~cluster or field membership,
	(6)~inclination ($90\degr \equiv$ edge-on),
	(7)~internal extinction correction (including 0.27 mag of face-on extinction),
	(8)~absolute rest-frame $B$-band magnitude (for $\Omega_{\Lambda} = 0.7$,
	$\Omega_{m} =0.3$, $H_0 = 70$ \kms~Mpc\per{} cosmology),
	(9)~rotation velocity,
	(10)~weight in TFR fits.
	}%
\centering%
\begin{tabular}{lrrccccccc}
\hline
\hspace{\stretch{1}} ID \hspace{\stretch{1}} & \hspace{\stretch{1}} R.A. \hspace{\stretch{1}} &%
\hspace{\stretch{1}} Dec. \hspace{\stretch{1}} & $z$ & Mem. & $i$ &%
$A^{\rmn{i}}$ & $M_B$ & $\log{\vrot}$ & $w_\rmn{TF}$ \\
  & \hspace{\stretch{1}} [J2000] \hspace{\stretch{1}} &%
\hspace{\stretch{1}} [J2000] \hspace{\stretch{1}} &  &  & [deg] &%
[mag] & [mag] & [dex] & \\
\hline
    MS0440\_101 &      04 43 14.6 &        02 05 49 & $   0.819$ &   F & $  66$ & $    0.59\pm  0.10$ & $  -21.72\pm  0.28$ & $    2.23$\asymerr{  0.05}{  0.03} &  0.017     \\
    MS0440\_140 &      04 43 09.0 &        02 06 13 & $   0.316$ &   F & $  55$ & $    0.45\pm  0.05$ & $  -19.58\pm  0.21$ & $    2.06$\asymerr{  0.07}{  0.06} &  0.014     \\
    MS0440\_188 &      04 43 19.3 &        02 06 42 & $   0.491$ &   F & $  53$ & $    0.43\pm  0.05$ & $  -20.91\pm  0.19$ & $    2.02$\asymerr{  0.21}{  0.20} &  0.004     \\
    MS0440\_207 &      04 43 16.8 &        02 07 01 & $   0.087$ &   F & $  51$ & $    0.41\pm  0.08$ & $  -18.79\pm  0.30$ & $    1.99$\asymerr{  0.06}{  0.06} &  \noweight \\
    MS0440\_273 &      04 43 16.4 &        02 07 31 & $   0.283$ &   F & $  79$ & $    0.96\pm  0.10$ & $  -20.10\pm  0.23$ & $    2.02$\asymerr{  0.04}{  0.03} &  0.018     \\
    MS0440\_311 &      04 43 06.2 &        02 07 51 & $   0.470$ &   F & $  39$ & $    0.34\pm  0.02$ & $  -21.55\pm  0.18$ & $    1.96$\asymerr{  0.10}{  0.09} &  0.011     \\
    MS0440\_319 &      04 43 07.8 &        02 08 06 & $   0.138$ &   F & $  82$ & $    0.96\pm  0.07$ & $  -19.19\pm  0.38$ & $    1.98$\asymerr{  0.03}{  0.03} &  \noweight \\
    MS0440\_538 &      04 43 18.1 &        02 09 43 & $   0.213$ &   F & $  59$ & $    0.49\pm  0.07$ & $  -21.87\pm  0.32$ & $    2.39$\asymerr{  0.03}{  0.02} &  \noweight \\
    MS0440\_616 &      04 42 58.8 &        02 09 37 & $   0.211$ &   F & $  76$ & $    0.84\pm  0.14$ & $  -20.00\pm  0.34$ & $    2.11$\asymerr{  0.05}{  0.05} &  \noweight \\
    MS0440\_627 &      04 43 14.9 &        02 09 33 & $   0.265$ &   F & $  74$ & $    0.76\pm  0.17$ & $  -20.49\pm  0.32$ & $    2.07$\asymerr{  0.02}{  0.02} &  0.019     \\
    MS0440\_635 &      04 43 27.2 &        02 09 43 & $   0.237$ &   F & $  80$ & $    0.96\pm  0.10$ & $  -20.23\pm  0.24$ & $    2.16$\asymerr{  0.01}{  0.01} &  \noweight \\
    MS0440\_657 &      04 42 50.5 &        02 09 45 & $   0.265$ &   F & $  40$ & $    0.34\pm  0.03$ & $  -19.94\pm  0.21$ & $    2.18$\asymerr{  0.07}{  0.06} &  0.014     \\
    MS0440\_735 &      04 43 07.2 &        02 10 21 & $   0.181$ &   F & $  79$ & $    0.95\pm  0.04$ & $  -20.53\pm  0.33$ & $    2.26$\asymerr{  0.01}{  0.01} &  \noweight \\
    MS0440\_849 &      04 43 14.4 &        02 10 30 & $   0.401$ &   F & $  67$ & $    0.60\pm  0.11$ & $  -20.39\pm  0.22$ & $    2.06$\asymerr{  0.05}{  0.05} &  0.016     \\
   MS0440\_1109 &      04 43 05.9 &        02 10 45 & $   0.239$ &   F & $  49$ & $    0.39\pm  0.01$ & $  -19.06\pm  0.29$ & $    2.08$\asymerr{  0.04}{  0.04} &  \noweight \\
   MS0440\_1131 &      04 43 06.7 &        02 12 15 & $   0.318$ &   F & $  67$ & $    0.60\pm  0.10$ & $  -21.08\pm  0.26$ & $    2.23$\asymerr{  0.03}{  0.03} &  0.018     \\
   MS0440\_1157 &      04 43 10.9 &        02 11 32 & $   0.401$ &   F & $  48$ & $    0.39\pm  0.04$ & $  -20.40\pm  0.18$ & $    2.25$\asymerr{  0.05}{  0.04} &  0.017     \\
     AC114\_115 &      22 58 59.7 &       -34 50 52 & $   0.500$ &   F & $  65$ & $    0.57\pm  0.09$ & $  -21.58\pm  0.10$ & $    2.22$\asymerr{  0.02}{  0.02} &  0.020     \\
     AC114\_264 &      22 58 55.1 &       -34 49 49 & $   0.098$ &   F & $  58$ & $    0.48\pm  0.02$ & $  -17.64\pm  0.07$ & $    1.95$\asymerr{  0.08}{  0.06} &  \noweight \\
     AC114\_391 &      22 58 45.6 &       -34 49 03 & $   0.567$ &   F & $  59$ & $    0.48\pm  0.02$ & $  -21.72\pm  0.03$ & $    2.25$\asymerr{  0.01}{  0.02} &  0.021     \\
     AC114\_553 &      22 58 56.7 &       -34 48 18 & $   0.210$ &   F & $  71$ & $    0.68\pm  0.14$ & $  -20.73\pm  0.20$ & $    2.38$\asymerr{  0.01}{  0.01} &  \noweight \\
     AC114\_700 &      22 58 33.7 &       -34 47 43 & $   0.351$ &   F & $  60$ & $    0.49\pm  0.02$ & $  -20.20\pm  0.06$ & $    2.23$\asymerr{  0.04}{  0.04} &  0.019     \\
     AC114\_810 &      22 58 46.1 &       -34 46 00 & $   0.354$ &   F & $  39$ & $    0.34\pm  0.03$ & $  -20.27\pm  0.06$ & $    2.25$\asymerr{  0.08}{  0.07} &  0.014     \\
     AC114\_875 &      22 58 51.2 &       -34 46 21 & $   0.171$ &   F & $  67$ & $    0.59\pm  0.03$ & $  -19.29\pm  0.12$ & $    2.00$\asymerr{  0.03}{  0.03} &  \noweight \\
       A370\_39 &      02 39 48.1 &       -01 38 16 & $   0.325$ &   F & $  61$ & $    0.51\pm  0.13$ & $  -19.08\pm  0.17$ & $    2.00$\asymerr{  0.05}{  0.05} &  \noweight \\
      A370\_119 &      02 40 02.5 &       -01 37 13 & $   0.564$ &   F & $  66$ & $    0.58\pm  0.03$ & $  -21.83\pm  0.10$ & $    2.38$\asymerr{  0.03}{  0.04} &  0.019     \\
      A370\_157 &      02 39 55.5 &       -01 36 59 & $   0.542$ &   F & $  69$ & $    0.64\pm  0.12$ & $  -20.37\pm  0.17$ & $    2.19$\asymerr{  0.04}{  0.04} &  0.018     \\
      A370\_183 &      02 40 00.4 &       -01 36 38 & $   0.361$ &   F & $  49$ & $    0.40\pm  0.05$ & $  -20.81\pm  0.11$ & $    2.40$\asymerr{  0.05}{  0.05} &  0.017     \\
      A370\_210 &      02 40 00.9 &       -01 36 16 & $   0.230$ &   F & $  55$ & $    0.44\pm  0.02$ & $  -20.44\pm  0.03$ & $    1.69$\asymerr{  0.06}{  0.05} &  \noweight \\
      A370\_292 &      02 39 57.8 &       -01 35 49 & $   0.542$ &   F & $  72$ & $    0.72\pm  0.05$ & $  -20.13\pm  0.09$ & $    1.84$\asymerr{  0.08}{  0.07} &  0.013     \\
      A370\_319 &      02 39 51.8 &       -01 35 21 & $   0.305$ &   F & $  46$ & $    0.38\pm  0.01$ & $  -21.01\pm  0.19$ & $    1.99$\asymerr{  0.08}{  0.15} &  0.008     \\
      A370\_401 &      02 39 54.5 &       -01 35 04 & $   0.346$ &   F & $  50$ & $    0.41\pm  0.01$ & $  -20.97\pm  0.15$ & $    2.26$\asymerr{  0.04}{  0.04} &  0.018     \\
      A370\_406 &      02 40 13.9 &       -01 35 05 & $   0.571$ &   F & $  49$ & $    0.40\pm  0.07$ & $  -21.83\pm  0.12$ & $    2.22$\asymerr{  0.07}{  0.06} &  0.015     \\
      A370\_540 &      02 40 09.7 &       -01 32 05 & $   0.173$ &   F & $  68$ & $    0.63\pm  0.12$ & $  -18.12\pm  0.21$ & $    1.66$\asymerr{  0.04}{  0.04} &  \noweight \\
      A370\_582 &      02 40 10.5 &       -01 33 29 & $   0.207$ &   F & $  77$ & $    0.88\pm  0.13$ & $  -19.96\pm  0.21$ & $    2.21$\asymerr{  0.04}{  0.04} &  \noweight \\
      A370\_620 &      02 40 04.3 &       -01 32 59 & $   0.250$ &   F & $  65$ & $    0.57\pm  0.09$ & $  -21.11\pm  0.17$ & $    2.22$\asymerr{  0.02}{  0.02} &  0.020     \\
      A370\_630 &      02 39 50.3 &       -01 34 22 & $   0.225$ &   F & $  75$ & $    0.78\pm  0.06$ & $  -19.01\pm  0.26$ & $    1.99$\asymerr{  0.01}{  0.01} &  \noweight \\
      A370\_650 &      02 39 57.8 &       -01 33 10 & $   0.547$ &   F & $  40$ & $    0.35\pm  0.03$ & $  -21.47\pm  0.11$ & $    2.37$\asymerr{  0.08}{  0.06} &  0.014     \\
      A370\_751 &      02 39 57.5 &       -01 34 32 & $   0.256$ &   F & $  56$ & $    0.45\pm  0.02$ & $  -20.07\pm  0.23$ & $    2.13$\asymerr{  0.01}{  0.01} &  0.020     \\
     CL0054\_62 &      00 57 00.9 &       -27 44 18 & $   0.537$ &   F & $  42$ & $    0.36\pm  0.03$ & $  -21.91\pm  0.09$ & $    2.36$\asymerr{  0.05}{  0.05} &  0.017     \\
     CL0054\_83 &      00 56 52.4 &       -27 44 06 & $   0.718$ &   F & $  75$ & $    0.80\pm  0.16$ & $  -21.97\pm  0.22$ & $    2.32$\asymerr{  0.02}{  0.01} &  0.020     \\
     CL0054\_89 &      00 56 52.4 &       -27 44 04 & $   0.537$ &   F & $  53$ & $    0.43\pm  0.05$ & $  -20.54\pm  0.10$ & $    2.19$\asymerr{  0.06}{  0.05} &  0.017     \\
    CL0054\_126 &      00 56 59.3 &       -27 43 41 & $   0.237$ &   F & $  77$ & $    0.87\pm  0.13$ & $  -18.14\pm  0.21$ & $    1.49$\asymerr{  0.07}{  0.07} &  \noweight \\
    CL0054\_137 &      00 56 55.7 &       -27 43 25 & $   0.297$ &   F & $  43$ & $    0.36\pm  0.03$ & $  -19.99\pm  0.13$ & $    2.19$\asymerr{  0.06}{  0.06} &  0.016     \\
    CL0054\_138 &      00 56 46.3 &       -27 42 50 & $   0.237$ &   F & $  84$ & $    0.96\pm  0.09$ & $  -21.84\pm  0.18$ & $    2.36$\asymerr{  0.02}{  0.02} &  \noweight \\
    CL0054\_284 &      00 56 55.7 &       -27 42 17 & $   0.815$ &   F & $  50$ & $    0.41\pm  0.04$ & $  -21.30\pm  0.20$ & $    2.08$\asymerr{  0.05}{  0.04} &  0.017     \\
    CL0054\_354 &      00 57 02.3 &       -27 41 31 & $   0.224$ &   F & $  46$ & $    0.38\pm  0.03$ & $  -20.48\pm  0.05$ & $    1.91$\asymerr{  0.04}{  0.04} &  \noweight \\
    CL0054\_407 &      00 57 00.7 &       -27 41 05 & $   0.275$ &   F & $  38$ & $    0.34\pm  0.01$ & $  -19.88\pm  0.05$ & $    2.32$\asymerr{  0.07}{  0.06} &  0.015     \\
    CL0054\_454 &      00 57 09.9 &       -27 40 44 & $   0.298$ &   F & $  73$ & $    0.73\pm  0.16$ & $  -21.69\pm  0.20$ & $    2.38$\asymerr{  0.02}{  0.02} &  0.020     \\
    CL0054\_579 &      00 57 12.3 &       -27 40 14 & $   0.577$ &   F & $  34$ & $    0.32\pm  0.02$ & $  -22.42\pm  0.10$ & $    2.18$\asymerr{  0.07}{  0.06} &  0.015     \\
    CL0054\_588 &      00 57 09.1 &       -27 40 27 & $   0.911$ &   F & $  29$ & $    0.31\pm  0.02$ & $  -21.72\pm  0.25$ & $    2.28$\asymerr{  0.08}{  0.07} &  0.013     \\
    CL0054\_686 &      00 56 50.9 &       -27 38 01 & $   0.710$ &   F & $  44$ & $    0.37\pm  0.06$ & $  -21.68\pm  0.16$ & $    2.25$\asymerr{  0.11}{  0.09} &  0.010     \\
    CL0054\_688 &      00 56 53.8 &       -27 37 14 & $   0.298$ &   F & $  65$ & $    0.57\pm  0.09$ & $  -19.62\pm  0.16$ & $    1.96$\asymerr{  0.02}{  0.02} &  0.020     \\
    CL0054\_779 &      00 57 14.5 &       -27 38 04 & $   1.004$ &   F & $  38$ & $    0.34\pm  0.02$ & $  -22.40\pm  0.29$ & $    2.19$\asymerr{  0.06}{  0.05} &  0.015     \\
    CL0054\_803 &      00 56 57.6 &       -27 38 24 & $   0.162$ &   F & $  50$ & $    0.40\pm  0.04$ & $  -19.40\pm  0.19$ & $    1.96$\asymerr{  0.04}{  0.03} &  \noweight \\
    CL0054\_827 &      00 57 09.1 &       -27 38 36 & $   0.583$ &   F & $  60$ & $    0.49\pm  0.07$ & $  -21.08\pm  0.13$ & $    2.19$\asymerr{  0.04}{  0.04} &  0.019     \\
    CL0054\_892 &      00 56 55.6 &       -27 39 08 & $   0.585$ &   F & $  32$ & $    0.32\pm  0.03$ & $  -22.11\pm  0.12$ & $    2.39$\asymerr{  0.14}{  0.10} &  0.008     \\
    CL0054\_927 &      00 57 07.9 &       -27 39 28 & $   0.653$ &   F & $  39$ & $    0.34\pm  0.02$ & $  -21.20\pm  0.13$ & $    2.19$\asymerr{  0.06}{  0.05} &  0.016     \\
    CL0054\_937 &      00 56 56.8 &       -27 39 34 & $   0.603$ &   F & $  65$ & $    0.57\pm  0.03$ & $  -21.10\pm  0.12$ & $    2.22$\asymerr{  0.05}{  0.08} &  0.015     \\
    CL0054\_979 &      00 56 47.2 &       -27 38 33 & $   0.660$ &   F & $  57$ & $    0.47\pm  0.06$ & $  -21.18\pm  0.14$ & $    2.10$\asymerr{  0.03}{  0.03} &  0.019     \\
    CL0054\_993 &      00 57 11.4 &       -27 39 46 & $   0.214$ &   F & $  70$ & $    0.66\pm  0.13$ & $  -21.13\pm  0.20$ & $    1.96$\asymerr{  0.02}{  0.02} &  \noweight \\
\hline
\end{tabular}
\end{minipage}
\end{table*}

\begin{table*}%
\begin{minipage}{0.88\textwidth}%
\contcaption{}%
\centering%
\begin{tabular}{lrrccccccccc}
\hline
\hspace{\stretch{1}} ID \hspace{\stretch{1}} & \hspace{\stretch{1}} R.A. \hspace{\stretch{1}} &%
\hspace{\stretch{1}} Dec. \hspace{\stretch{1}} & $z$ & Mem. & $i$ &%
$A^{\rmn{i}}$ & $M_B$ & $\log{\vrot}$ & $w_\rmn{TF}$ \\
  & \hspace{\stretch{1}} [J2000] \hspace{\stretch{1}} &%
\hspace{\stretch{1}} [J2000] \hspace{\stretch{1}} &  &  & [deg] &%
[mag] & [mag] & [dex] & \\
\hline
   CL0054\_1011 &      00 56 58.9 &       -27 40 20 & $   0.171$ &   F & $  66$ & $    0.58\pm  0.03$ & $  -20.21\pm  0.07$ & $    2.21$\asymerr{  0.01}{  0.01} &  \noweight \\
   CL0054\_1054 &      00 56 59.8 &       -27 38 08 & $   0.830$ &   F & $  67$ & $    0.61\pm  0.11$ & $  -21.48\pm  0.24$ & $    2.26$\asymerr{  0.04}{  0.03} &  0.018     \\
     MS2053\_86 &      20 56 12.4 &       -04 38 26 & $   0.196$ &   F & $  74$ & $    0.78\pm  0.16$ & $  -19.60\pm  0.23$ & $    2.11$\asymerr{  0.01}{  0.01} &  \noweight \\
    MS2053\_371 &      20 56 17.8 &       -04 38 01 & $   0.521$ &   F & $  55$ & $    0.44\pm  0.02$ & $  -21.15\pm  0.07$ & $    2.15$\asymerr{  0.05}{  0.05} &  0.017     \\
    MS2053\_404 &      20 56 18.7 &       -04 37 07 & $   0.384$ &   F & $  69$ & $    0.65\pm  0.04$ & $  -21.74\pm  0.05$ & $    2.45$\asymerr{  0.01}{  0.01} &  0.022     \\
    MS2053\_435 &      20 56 18.9 &       -04 40 04 & $   0.520$ &   F & $  53$ & $    0.43\pm  0.02$ & $  -21.83\pm  0.14$ & $    2.33$\asymerr{  0.01}{  0.02} &  0.021     \\
    MS2053\_455 &      20 56 20.0 &       -04 35 54 & $   0.174$ &   F & $  79$ & $    0.95\pm  0.04$ & $  -20.09\pm  0.18$ & $    2.15$\asymerr{  0.02}{  0.02} &  \noweight \\
    MS2053\_470 &      20 56 19.5 &       -04 38 47 & $   0.371$ &   F & $  77$ & $    0.88\pm  0.07$ & $  -20.69\pm  0.08$ & $    2.20$\asymerr{  0.03}{  0.03} &  0.020     \\
    MS2053\_741 &      20 56 23.2 &       -04 34 41 & $   0.335$ &   F & $  77$ & $    0.85\pm  0.07$ & $  -20.41\pm  0.07$ & $    1.94$\asymerr{  0.04}{  0.04} &  0.019     \\
    MS2053\_856 &      20 56 24.8 &       -04 35 34 & $   0.261$ &   F & $  70$ & $    0.66\pm  0.04$ & $  -21.38\pm  0.06$ & $    2.29$\asymerr{  0.02}{  0.02} &  0.021     \\
    MS2053\_998 &      20 56 22.6 &       -04 41 32 & $   0.196$ &   F & $  72$ & $    0.71\pm  0.05$ & $  -19.99\pm  0.18$ & $    2.10$\asymerr{  0.02}{  0.02} &  \noweight \\
   MS2053\_1105 &      20 56 29.6 &       -04 38 08 & $   0.408$ &   F & $  77$ & $    0.85\pm  0.07$ & $  -19.49\pm  0.09$ & $    1.83$\asymerr{  0.03}{  0.03} &  \noweight \\
   MS2053\_1296 &      20 56 34.2 &       -04 38 02 & $   0.058$ &   F & $  57$ & $    0.46\pm  0.06$ & $  -15.03\pm  0.26$ & $    1.45$\asymerr{  0.16}{  0.12} &  \noweight \\
    MS1054\_F02 &      10 56 48.3 &       -03 37 33 & $   0.180$ &   F & $  66$ & $    0.58\pm  0.03$ & $  -19.83\pm  0.11$ & $    2.00$\asymerr{  0.01}{  0.01} &  \noweight \\
    MS1054\_F04 &      10 56 56.0 &       -03 37 28 & $   0.230$ &   F & $  80$ & $    0.96\pm  0.03$ & $  -19.04\pm  0.06$ & $    1.73$\asymerr{  0.07}{  0.09} &  \noweight \\
    MS1054\_F05 &      10 57 01.3 &       -03 35 44 & $   0.249$ &   F & $  79$ & $    0.96\pm  0.04$ & $  -19.06\pm  0.08$ & $    2.06$\asymerr{  0.03}{  0.02} &  \noweight \\
    MS1054\_F06 &      10 56 53.0 &       -03 38 41 & $   0.259$ &   F & $  78$ & $    0.91\pm  0.06$ & $  -20.44\pm  0.08$ & $    2.07$\asymerr{  0.01}{  0.02} &  0.021     \\
    MS1054\_F08 &      10 57 08.2 &       -03 37 34 & $   0.287$ &   F & $  69$ & $    0.65\pm  0.04$ & $  -19.47\pm  0.06$ & $    2.02$\asymerr{  0.03}{  0.03} &  \noweight \\
    MS1054\_F10 &      10 57 12.3 &       -03 37 17 & $   0.324$ &   F & $  68$ & $    0.63\pm  0.04$ & $  -19.88\pm  0.05$ & $    1.99$\asymerr{  0.04}{  0.03} &  0.019     \\
    MS1054\_F11 &      10 57 08.2 &       -03 36 42 & $   0.325$ &   F & $  45$ & $    0.37\pm  0.01$ & $  -19.58\pm  0.02$ & $    2.02$\asymerr{  0.02}{  0.02} &  0.021     \\
    MS1054\_F12 &      10 57 11.5 &       -03 36 44 & $   0.325$ &   F & $  64$ & $    0.55\pm  0.03$ & $  -20.99\pm  0.04$ & $    2.42$\asymerr{  0.02}{  0.02} &  0.021     \\
    MS1054\_F14 &      10 56 54.7 &       -03 39 00 & $   0.429$ &   F & $  77$ & $    0.86\pm  0.07$ & $  -18.75\pm  0.09$ & $    1.86$\asymerr{  0.05}{  0.04} &  \noweight \\
    MS1054\_F16 &      10 57 01.2 &       -03 34 20 & $   0.470$ &   F & $  38$ & $    0.34\pm  0.01$ & $  -21.62\pm  0.04$ & $    2.26$\asymerr{  0.02}{  0.02} &  0.021     \\
    MS1054\_F18 &      10 57 03.7 &       -03 38 33 & $   0.553$ &   F & $  58$ & $    0.47\pm  0.02$ & $  -20.69\pm  0.06$ & $    2.17$\asymerr{  0.02}{  0.02} &  0.021     \\
    MS1054\_F19 &      10 56 50.7 &       -03 35 39 & $   0.684$ &   F & $  76$ & $    0.84\pm  0.07$ & $  -20.92\pm  0.08$ & $    2.19$\asymerr{  0.04}{  0.02} &  0.019     \\
    MS1054\_F20 &      10 57 05.7 &       -03 36 26 & $   0.686$ &   F & $  81$ & $    0.96\pm  0.01$ & $  -20.95\pm  0.06$ & $    2.12$\asymerr{  0.03}{  0.02} &  0.020     \\
    MS1054\_F21 &      10 56 48.6 &       -03 35 42 & $   0.756$ &   F & $  50$ & $    0.41\pm  0.01$ & $  -21.18\pm  0.04$ & $    2.27$\asymerr{  0.04}{  0.03} &  0.019     \\
    MS1054\_F22 &      10 57 07.8 &       -03 37 04 & $   0.896$ &   F & $  69$ & $    0.63\pm  0.04$ & $  -22.48\pm  0.06$ & $    2.38$\asymerr{  0.01}{  0.01} &  0.022     \\
      AC114\_18 &      22 58 48.5 &       -34 51 39 & $   0.306$ &   C & $  72$ & $    0.72\pm  0.22$ & $  -21.44\pm  0.23$ & $    2.16$\asymerr{  0.02}{  0.02} &  0.048     \\
     AC114\_142 &      22 58 52.0 &       -34 50 42 & $   0.325$ &   C & $  43$ & $    0.36\pm  0.03$ & $  -19.92\pm  0.05$ & $    2.07$\asymerr{  0.05}{  0.04} &  0.047     \\
     AC114\_193 &      22 58 58.9 &       -34 50 20 & $   0.307$ &   C & $  78$ & $    0.89\pm  0.07$ & $  -21.18\pm  0.08$ & $    2.05$\asymerr{  0.01}{  0.01} &  0.053     \\
     AC114\_768 &      22 58 35.8 &       -34 45 47 & $   0.314$ &   C & $  59$ & $    0.49\pm  0.07$ & $  -22.28\pm  0.10$ & $    2.41$\asymerr{  0.04}{  0.03} &  0.049     \\
     AC114\_930 &      22 58 34.0 &       -34 46 52 & $   0.306$ &   C & $  60$ & $    0.49\pm  0.02$ & $  -20.72\pm  0.08$ & $    1.91$\asymerr{  0.14}{  0.08} &  0.028     \\
     AC114\_959 &      22 58 49.3 &       -34 47 01 & $   0.313$ &   C & $  52$ & $    0.42\pm  0.02$ & $  -21.26\pm  0.08$ & $    1.93$\asymerr{  0.08}{  0.06} &  0.039     \\
    AC114\_1001 &      22 58 30.1 &       -34 47 21 & $   0.307$ &   C & $  50$ & $    0.40\pm  0.04$ & $  -21.02\pm  0.07$ & $    1.94$\asymerr{  0.08}{  0.09} &  0.036     \\
      A370\_532 &      02 39 51.0 &       -01 32 12 & $   0.374$ &   C & $  82$ & $    0.96\pm  0.12$ & $  -22.31\pm  0.18$ & $    2.27$\asymerr{  0.01}{  0.01} &  0.051     \\
      A370\_538 &      02 39 58.2 &       -01 32 32 & $   0.373$ &   C & $  68$ & $    0.62\pm  0.11$ & $  -21.64\pm  0.17$ & $    2.07$\asymerr{  0.02}{  0.02} &  0.050     \\
      A370\_555 &      02 39 46.4 &       -01 32 17 & $   0.378$ &   C & $  71$ & $    0.69\pm  0.14$ & $  -21.13\pm  0.17$ & $    2.39$\asymerr{  0.02}{  0.02} &  0.050     \\
    CL0054\_358 &      00 57 04.5 &       -27 41 30 & $   0.564$ &   C & $  41$ & $    0.35\pm  0.03$ & $  -21.70\pm  0.10$ & $    2.06$\asymerr{  0.09}{  0.08} &  0.037     \\
    CL0054\_609 &      00 56 45.2 &       -27 38 08 & $   0.558$ &   C & $  59$ & $    0.49\pm  0.07$ & $  -21.52\pm  0.12$ & $    1.98$\asymerr{  0.04}{  0.03} &  0.049     \\
    CL0054\_643 &      00 56 45.4 &       -27 38 06 & $   0.558$ &   C & $  51$ & $    0.41\pm  0.08$ & $  -20.72\pm  0.13$ & $    2.12$\asymerr{  0.08}{  0.06} &  0.040     \\
    CL0054\_714 &      00 56 46.2 &       -27 37 23 & $   0.562$ &   C & $  45$ & $    0.37\pm  0.03$ & $  -20.46\pm  0.11$ & $    2.00$\asymerr{  0.07}{  0.05} &  0.044     \\
    CL0054\_725 &      00 56 44.8 &       -27 37 46 & $   0.557$ &   C & $  63$ & $    0.53\pm  0.08$ & $  -21.64\pm  0.13$ & $    2.33$\asymerr{  0.02}{  0.02} &  0.051     \\
    CL0054\_799 &      00 56 56.4 &       -27 38 22 & $   0.554$ &   C & $  60$ & $    0.50\pm  0.07$ & $  -22.40\pm  0.12$ & $    2.51$\asymerr{  0.04}{  0.03} &  0.049     \\
    CL0054\_860 &      00 56 49.6 &       -27 38 51 & $   0.559$ &   C & $  53$ & $    0.43\pm  0.05$ & $  -21.38\pm  0.11$ & $    2.13$\asymerr{  0.04}{  0.03} &  0.049     \\
    CL0054\_918 &      00 57 05.5 &       -27 40 04 & $   0.557$ &   C & $  47$ & $    0.38\pm  0.04$ & $  -23.07\pm  0.11$ & $    2.38$\asymerr{  0.06}{  0.05} &  0.044     \\
    CL0054\_966 &      00 56 48.4 &       -27 40 03 & $   0.559$ &   C & $  57$ & $    0.46\pm  0.06$ & $  -21.06\pm  0.12$ & $    2.29$\asymerr{  0.09}{  0.07} &  0.037     \\
    MS1054\_C01 &      10 57 12.0 &       -03 36 50 & $   0.828$ &   C & $  80$ & $    0.96\pm  0.03$ & $  -21.29\pm  0.05$ & $    2.10$\asymerr{  0.03}{  0.03} &  0.050     \\
   MS1054\_1403 &      10 57 03.8 &       -03 37 43 & $   0.813$ &   C & $  70$ & $    0.66\pm  0.04$ & $  -23.02\pm  0.05$ & $    2.45$\asymerr{  0.01}{  0.01} &  0.053     \\
   MS1054\_2011 &      10 57 07.1 &       -03 35 40 & $   0.841$ &   C & $  53$ & $    0.42\pm  0.02$ & $  -20.86\pm  0.03$ & $    2.05$\asymerr{  0.05}{  0.03} &  0.048     \\
\hline
\end{tabular}
\end{minipage}
\end{table*}

\vspace{-1.2\baselineskip}
\section{`Matched' cluster and field samples}
\label{sec:matched}

Cluster membership has been assigned based on redshifts and
velocity-dispersions from the literature
\citep{GM01,SMGMSWFH91,HFKvD02}.  These are given in Table
\ref{tab:clusters}, along with the corresponding $3\sigma$ redshift
limits we have adopted ($\Delta \zcl$).  In each field, galaxies with
$\zcl - \Delta \zcl \le z \le \zcl + \Delta \zcl$ are considered to be
cluster members.

\begin{figure}%
\centering%
\includegraphics[height=0.45\textwidth,angle=270]%
		{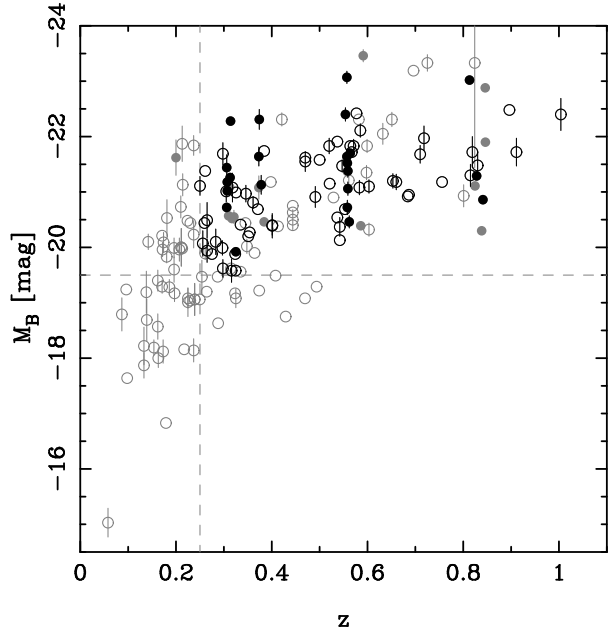}%
\caption{\label{fig:m_vs_z}%
	Absolute rest-frame $B$-band magnitude versus redshift for
	\emph{all} our field (open points) and cluster (filled points)
	galaxies with emission lines.
	Points in our final `matched' TFR sample are black, while those not 
	included are grey. Grey points thus denote galaxies
	for which no lines pass our quality criteria or which do not 
	meet our $M_B$ or $z$ cuts.
	Error bars are not shown where they are smaller than the symbols.
	Adopted cuts for the `matched' samples are shown by dashed lines.
}
\end{figure}

Our target selection has been performed in a way designed to give
easily comparable samples of field and cluster galaxies, however
further efforts are required to ensure these samples are well matched.
While the cluster galaxies are located at particular redshifts,
between $0.30 \la z \la 0.85$, the field galaxies span a much wider
range in redshift, and consequently in absolute magnitude.  As the TFR
may evolve with redshift, irrespective of environment
(e.g. see \papertwo), care must be
taken to avoid this complicating the comparison between cluster and
field.  In particular, the evolution of low-luminosity galaxies, which
are not represented in our sample at higher redshifts, is particularly
unconstrained.
We therefore impose cuts of $z \geq 0.25$ and $M_B \leq -19.5$ mag,
simply chosen to better match the distribution of field galaxies to
that of the cluster sample, as indicated in Fig.~\ref{fig:m_vs_z}.

The `matched' TFR sample used for the cluster$-$field comparison in
this paper thus contains a total of $80$ galaxies, comprising $58$
field and $22$ cluster galaxies.

Now we have established samples of field and cluster galaxies over
similar epochs and luminosity ranges, we can investigate whether the
samples differ in other ways.  The distributions of $M_B$, $\vrot$,
$\rdspec$ and $\rdphot$ are shown in Fig.~\ref{fig:param_distrib},
with the K-S test confidence levels that the parent distributions of
the two samples are the same.  Note from Figures
\ref{fig:param_distrib}(a) and \ref{fig:m_vs_z} that the cluster and
field galaxies cover a very similar range in $M_B$, with a hint that
the cluster galaxies extend to brighter magnitudes.  There should be
absolutely no difference in the selection of galaxies at the bright
end of this distribution.  This is therefore a first indication that
cluster spirals may be brighter than those in the field.

The distributions of galaxy size, in terms of both photometric and
spectroscopic scalelength (Fig.~\ref{fig:param_distrib} panels (b) and
(d)), are similar for the field and cluster samples. For $\rdphot$
they are practically identical, although it is worth noting that the
cluster members do extend to larger values.  On the other hand, the
cluster $\rdspec$ distribution is restricted to lower values than the
field.

The two samples cover the same range in $\vrot$, although there is
evidence that the cluster galaxies have a slightly broader
distribution, possibly more skewed to lower values.  This is
considered further in the discussion (\S\ref{sec:discussion}).
However, these differences are minor, and point at real characteristics
of the galaxy population, rather than a selection bias.

\vspace{-1.2\baselineskip}
\section{Cluster versus field TFR} \label{sec:analysis}

\begin{figure}%
\centering%
\includegraphics[height=0.45\textwidth,angle=270]%
		{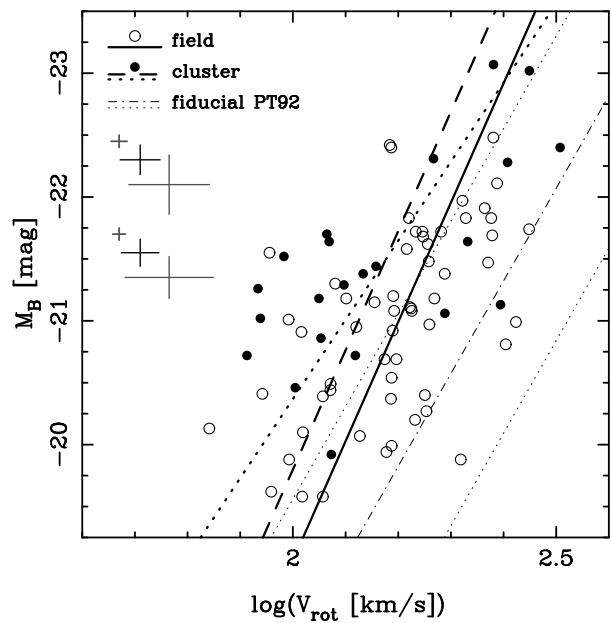}%
\caption{\label{fig:tf}%
	The Tully--Fisher relation for our `matched' samples of
	field (open points) and cluster (filled points) galaxies.
	The fiducial local relation of \citet{PT92} is marked by the thin 
	dot-dashed line, with its $3\sigma$ scatter delimited by thin dotted
	lines.  Weighted least-squares fits to the matched field sample 
	(solid line) and cluster sample (constrained to the field slope: 
	dashed line, free slope: dotted line) are also marked.
	The two sets of error bars shown on the left indicate the 
	10th-, 50th- and 90th-percentile errors for field (top) and 
	cluster (bottom) points.}
\end{figure}

Fig.~\ref{fig:tf} shows the TFR for our `matched' samples of field and
cluster galaxies.  A fiducial local field TFR is indicated by the thin
lines.  This is derived from the TFR of \citet[hereafter PT92]{PT92},
with a zero-point adjustment because PT92, while otherwise using the
internal extinction correction of \citet{TF85}, do not include the
$0.27$ mag of face-on extinction that is applied to our data.  The
fiducial PT92 TFR, adapted to our internal extinction correction, is
thus:
\begin{equation} \label{eqn:PT92}
M_{B}^{\rmn{PT92}}(\vrot) = -7.48\log{\vrot} - 3.37.
\end{equation}

The thick solid line in Fig.~\ref{fig:tf} is a fit to the matched field
sample.  This is a weighted, least-squares fit, minimising the
residuals in $\vrot$ (referred to as an `inverse' TFR fit) and
incorporating an intrinsic scatter term, as described in more detail
in \papertwo.  The thick dotted and dashed lines are fits to the
cluster sample, performed by the same method, except for the dashed
line the slope was fixed to that of the fit to the matched field sample.
The fit to the matched field sample, and cluster samples with free
and fixed slopes, respectively give
\begin{eqnarray} \label{eqn:tf_fits}
M_{B} &=& (-9.6\pm1.7) \cdot \log{\vrot} \:+\: (0.2\pm3.8)\; ,\nonumber\\
M_{B} &=& (-6.4\pm1.5) \cdot \log{\vrot} \:+\: (-7.6\pm5.5)\; ,\\
M_{B} &=& -9.6 \cdot \log{\vrot} \:+\: (-0.5\pm0.3)\; .\nonumber
\end{eqnarray}
The slope of the fit to the cluster sample is markedly shallower than
that to the matched field sample.  However, the slopes only differ
by $1.4\sigma$, so this not a significant result.

The weights ($w_\rmn{TF}$) assigned to each galaxy in the above fits
are given in Table \ref{tab:data}.  These are calculated from the
reciprocal of the sum of the squared uncertainties in $\vrot$ and $M_{B}$
and the intrinsic scatter.  The best fit is determined
iteratively, because of its dependence on the slope (used to convert the
$M_B$ error into one in $\vrot$) and intrinsic scatter.  The
$w_\rmn{TF}$ only differ by $\la 0.001$ between the two alternative
cluster fits; the values for the free-slope fit are given.

It can be seen that the field galaxies lie primarily along a
reasonably tight relation, with similar slope to the local fiducial
TFR, but with an offset to brighter magnitudes and/or lower rotation
velocities.  This is particularly clear when considering the full
field sample, unrestricted in $M_B$, as in \papertwo.  This overall,
systematic offset from the fiducial local TFR is of little concern for
this study.  As discussed more thoroughly in \papertwo, a comparison
with the intercept of the PT92 TFR must consider the different manners
in which the magnitudes and rotation velocities are measured for the
two studies.  
It is also likely that the absolute
calibration of the PT92 TFR is incorrect by $\unisim0.5$ mag,
as discussed further in \papertwo.
Correcting for this would bring the fiducial TFR into closer agreement
with our field sample, particularly for the low redshift objects.

The cluster members are preferentially located above the field
relation, particularly for galaxies with lower rotation velocities
$\la 150$ \kms, as also indicated by the shallower slope of their TFR
fit.  To help compare the field and cluster samples we can take
out the slope and examine the residuals from the fiducial TFR:
\begin{eqnarray} \label{eqn:dtf}
\dtf &=& M_{B} - M_{B}^{\rmn{PT92}}(\vrot) \nonumber \\
     &=& M_{B} - \left( -7.48\log{\vrot} - 3.37 \right) \; .
\end{eqnarray}

\begin{figure}
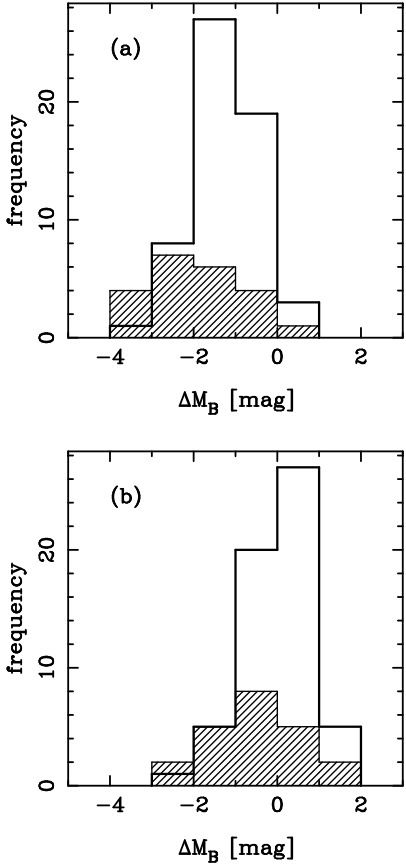
%
\centering%
\includegraphics[height=0.30\textwidth,angle=270]%
		{fig5a.eps}%
\bigskip\\%
\includegraphics[height=0.30\textwidth,angle=270]%
		{fig5b.eps}%
\caption{\label{fig:dtf_histo}%
	The offset from the fiducial 
	Tully--Fisher relation of PT92 for our `matched' samples of
	field (empty histogram) and cluster (shaded histogram)
	galaxies. Panel (a) shows the distributions of $\dtf$,
	while in (b) $\dtf$ has been corrected for the field evolution
	with redshift found in \papertwo.
	The K-S probabilities for the  
	cluster and field distributions being the same
	are (a) $0.1$\% and (b) $0.8$\%.
	}
\end{figure}

The difference between the cluster and field galaxies is particularly
evident in a histogram of $\dtf$, as shown in
Fig.~\ref{fig:dtf_histo}(a).  The peaks of the distributions are clearly
not aligned, such that cluster galaxies are generally brighter at a
fixed rotation velocity.  
A K-S test gives the probability of the
parent distributions being the same as $0.1$\%.

To assess this offset more quantitatively we can consider the mean and
variance of $\dtf$ for each of the samples.  These are calculated in a
similar manner to the TFR fitting method described in \papertwo.
Weighted means and variances are calculated, with weights assigned
from the measurement errors in combination with an
iteratively-determined intrinsic scatter.  The derived offset between
the cluster and field samples is $0.7\pm0.2$ mag. A $t$-test gives the
significance of this offset as $3\sigma$.

\begin{figure}%
\centering%
\includegraphics[height=0.45\textwidth,angle=270]%
		{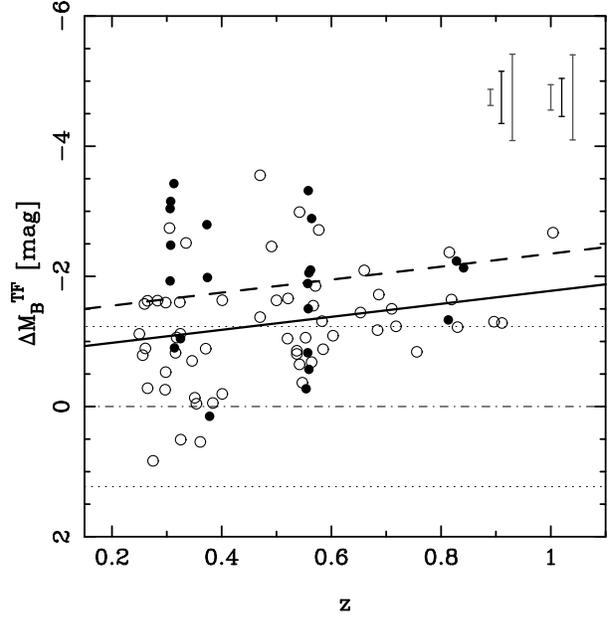}%
\caption{\label{fig:tf_res_z}%
	The residuals from the fiducial Tully--Fisher relation of PT92
	for our matched TFR samples of field (open points) and
	cluster (filled points) galaxies.
	The fiducial local relation of \citet{PT92} is marked by the thin 
	dot-dashed line, with its $3\sigma$ scatter delimited by thin dotted
	lines.  Weighted least-squares fits to the full field TFR sample 
	from \papertwo{} (solid line) 
	and the cluster sample (constrained to the field slope: dashed line) 
	are also marked.
	The two sets of error bars in the upper-right corner indicate the 
	10th-, 50th- and 90th-percentile errors for field (left) and
	cluster (right) points.}
\end{figure}

It could be suggested that the offset we find between field and
cluster galaxies is due to the combination of a general (field) trend
with redshift and a difference between the redshift distribution of
the field and cluster samples.  In order to demonstrate that this is
not the case we plot $\dtf$ versus redshift in
Fig.~\ref{fig:tf_res_z}.  Despite a possible trend in $\dtf$ with
redshift for the field population, as examined in \papertwo, the
offset is clearly still present, with cluster galaxies consistently
brighter for the same rotation velocity and redshift.  
The best-fitting field evolution from \papertwo{} is
\begin{equation} \label{eqn:field_dtf_z}
	\dtf = (-1.0\pm0.5) \cdot z + (0.8\pm0.2) \ \rmn{mag}
\end{equation}
Subtracting this field evolution does not change
either the size or significance of the cluster$-$field offset.  This
can also be seen in Fig.~\ref{fig:dtf_histo}(b), a histogram of $\dtf$
with the field evolution taken out, although the K-S significance declines
slightly.

Note that we have selected cluster galaxies in each field simply from
their redshifts coinciding with the targeted clusters.  It is possible
that a number of the galaxies we classify here as `field' actually
reside in separate high density regions.  Unfortunately, our sample is
not large enough to identify additional groups in our field. It will
be interesting to see if the field scatter is reduced in future
studies, when such groups can be excluded, and whether the group
spirals inhabit the same region of the TFR as our cluster sample.

\vspace{-1.2\baselineskip}
\section{Discussion} \label{sec:discussion}

\subsection{Origin of the TFR cluster$-$field offset}

It is clear that there is a significant difference between the cluster
and field galaxies in our sample.  
As the galaxies have all been selected, observed and analysed in the
same manner, it is very likely this difference is real.  Now we must
consider the reasons for this disparity.  The offset from the field
TFR may be due to some effect causing cluster galaxies to appear
brighter for a given rotation velocity, slower-rotating for a given
magnitude, or a combination of both.  Physically, both scenarios are
possible, an enhancement of star-formation would lead to a
brightening, while stripping of the dark matter halo by the cluster
potential could, at least hypothetically, decrease the galaxy mass and
hence lower $\vrot$.

Simulations by \citet[hereafter G03b]{G03b} find that $\vrot$ changes
little ($\la 5$\% decrease), even when over half of a galaxy's
dark-matter halo is stripped away.  This is because the halo is
truncated to a galactocentric radius that lies beyond the edge of the
luminous disk.  Within the region that the galaxy is luminous -- and
thus its rotation can be measured -- the halo is mostly unaffected,
the enclosed mass stays constant, and therefore $\vrot$ remains the
same.

G03b uses a pseudo-isothermal initial dark matter density
profile.  This is `cored' (finite at the centre), as opposed to the
`cuspy' halo profiles generally produced by CDM simulations
\citep{NFW96,M98}.  However, cored profiles \citep{vA85,B95} seem to
be required for galaxy haloes, from observations of individual
rotation curve shapes \citep[e.g.][]{GSKVK04}, and in order to solve
the problem of reproducing the TFR zeropoint in a hierarchical
universe \citep{NS3}.  If galaxy haloes are instead actually cuspy, it
would be even harder to remove dark matter from their inner regions
than found by G03b.  From this point of view their result
provides an upper limit on the feasible change in $\vrot$ due to tidal
interaction with a cluster.

However, higher resolution simulations using a different code, also
performed by G03b, find slightly larger decreases in $\vrot$ of
$\unisim 15$\%.  Adopting the slope of the PT92 TFR, this corresponds
to an apparent brightening at fixed rotation velocity of $\sim 0.5$
mag, comparable to the TFR offset we measure.

In addition, the G03b simulations discussed above are based on
galaxies with $\vrot = 250$ \kms.  Less massive galaxies, with
correspondingly less dense haloes, may well be more seriously
affected.  For example, G03 find that low surface brightness galaxies
with comparable masses, but much more extended haloes, suffer
decreases in $\vrot$ of $\ga 20$\%, while dwarf galaxies, with initial
$\vrot = 20$ \kms, are completely destroyed.  This could explain why
most of our cluster galaxies with large TFR offsets have low rotation
velocities, $\vrot \la 150$ \kms.  Further simulations would be helpful
to establish how the effectiveness of tidal stripping depends upon the 
initial rotation velocity of infalling galaxies.

With our present data, and the uncertainties concerning the dark
matter halo profile, we therefore cannot exclude tidal stripping of
the galaxy dark matter haloes as an origin for the cluster--field TFR
offset we measure.

The enhanced SFR hypothesis is supported by the increased
fraction of galaxies with strong E+A spectra (EW(H$\delta$) $\ga
5.0$\AA) in intermediate redshift clusters \citep{PSDCBBEO99,Tran03},
implying these galaxies have recently experienced a short star-burst
prior to truncation of their star-formation.  More direct evidence is
provided by a correlation between star-formation rate and offset from
the fiducial TFR, as suggested by our MS1054 sample in
\citet{MJthesis} and Milvang-Jensen \& Arag{\' o}n-Salamanca in
preparation, and which will be the subject of further study using our
entire TFR sample.

However, there may be a less straightforward reason why we observe
lower rotation velocities for cluster galaxies.  This could be a
symptom of cluster galaxies having rotation curves or emission surface
brightness profiles that are different from field galaxies.  Both of
these could cause a systematic divergence from the assumptions used in
\elfitpy, thereby affecting the measured value of $\vrot$.
\citet{Vogt04II} find spirals in local clusters with truncated
H$\alpha$ emission and deficient in HI, presumably due to removal of
gas from the outer regions of the disc through interactions with the
cluster environment. If spirals in our cluster sample are
significantly affected by this, then we may preferentially be
observing emission from nearer the centre of these galaxies.  This
could potentially bias our $\vrot$ measurements to lower values.
To look for any differences in the extent, quality and shape of the
rotation curves between the two samples, we can utilise the emission
line `traces' described in \S\ref{sec:spec}.

\begin{figure}%
\centering%
\includegraphics[height=0.35\textwidth,angle=270]%
		{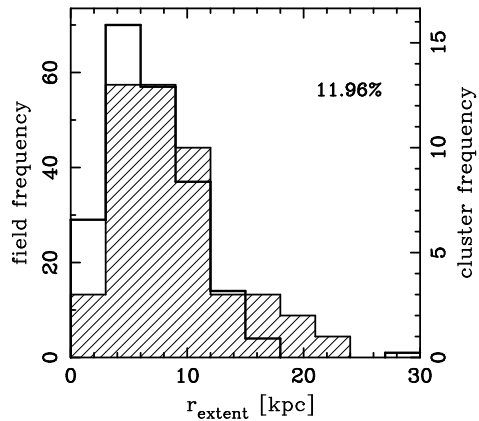}%
\caption{\label{fig:extent}%
	The $\rextent$ distribution for the lines used to measure
	$\vrot$ and $\rdspec$ of galaxies in the full TFR sample, in
	units of kpc.  The field (empty line) and cluster (hatched)
	histograms have been scaled to the same area.  The cluster
	galaxies have a similar distribution of extent compared with
	the field sample.  Note that the cluster galaxy emission lines
	can still often be traced out as far, or further, than for the
	field galaxies.
	The percentage in the top right corner indicates the
	confidence that the field and cluster samples are drawn from
	the same distribution, as given by a K-S test.
}
\end{figure}

Recall that $\rextent$ is the spatial distance, from the line centre,
to which we can reliably detect the emission above the background
noise.  This was determined by attempting to fit a Gaussian to the
emission in each spatial row, repeating the fit with different noise
realisations to determine average Gaussian parameters and their
uncertainties.  The sanity of these parameters and their significance,
as judged by the derived uncertainties, were then used to classify
each average fit as `good' or `bad', according to whether the
emission was reliably detected in that spatial row.  In addition,
isolated points, otherwise deemed to be `good', but separated from
other `good' points by more than two spatial pixels, were also judged
unreliable and hence classified as `bad' points.  The resultant values
of $\rextent$ are thus robust measurements of the extent to which the
emission lines can be reliably detected.
The distributions of emission line extent, $\rextent$, in units of
kpc, for all lines used to measure $\vrot$ for galaxies in the
`matched' samples, are shown in Fig.~\ref{fig:extent}.  It is clear
that there is very little difference between the extent of the
emission lines for cluster and field galaxies, and hence no evidence
of a bias that could affect the measured values of $\vrot$ and
$\rdspec$.  If anything, Fig.~\ref{fig:extent} suggests that we can
actually trace the emission out further in cluster galaxies than in
field galaxies.

\begin{figure}
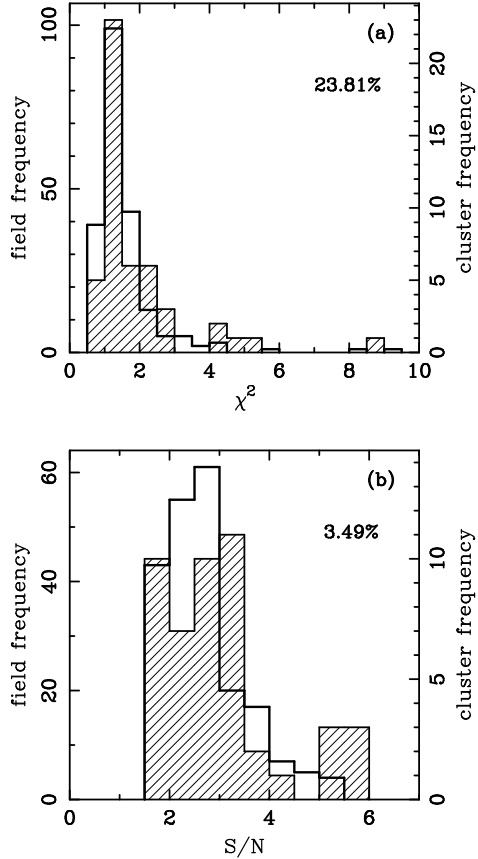
%
\centering%
\includegraphics[height=0.35\textwidth,angle=270]%
		{fig8a.eps}%
\bigskip\\%
\includegraphics[height=0.35\textwidth,angle=270]%
		{fig8b.eps}%
\caption{\label{fig:chi2sn}%
	The distributions of (a) $\chi_{\rmn{r}}^2$
	and (b) \snr for the lines used to measure 
	$\vrot$ and $\rdspec$ of galaxies in the full TFR sample.
	The field (empty line) and cluster (hatched) histograms have
	been scaled to the same area.
	There are no significant differences between
	the matched cluster and field samples.
	The percentage in the top right corner indicates the
	confidence that the field and cluster samples are drawn from
	the same distribution, as given by a K-S test.
}
\end{figure}

The distributions of the additional quality assessment quantities,
$\chi_{\rmn{r}}^2$ and \snr{} are shown in Fig.~\ref{fig:chi2sn}.
Again, there is no appreciable difference between the two samples,
apart from a hint that the lines of cluster galaxies extend to higher \snr{}
than those of the field galaxies.
We therefore conclude that there is no significant difference in the
extent or quality-of-fit of our cluster and field galaxy rotation
curves.

\begin{figure}
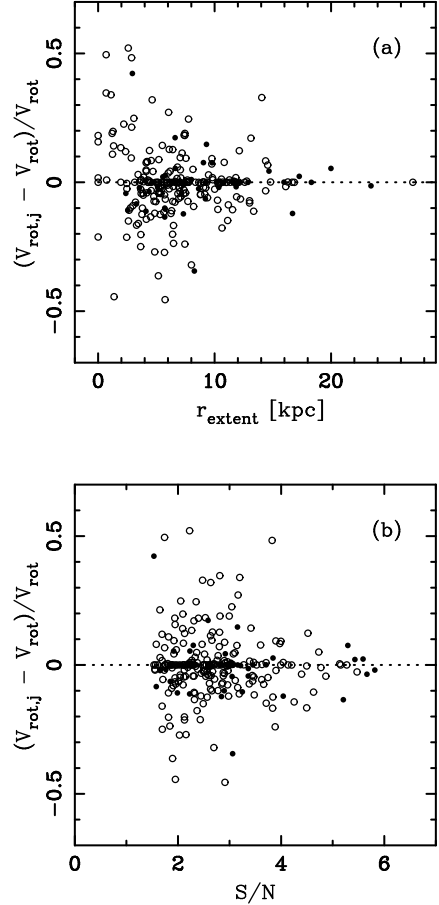
%
\centering%
\includegraphics[height=0.32\textwidth,angle=270]%
		{fig9a.eps}%
\bigskip\\%
\includegraphics[height=0.32\textwidth,angle=270]%
		{fig9b.eps}%
\caption{\label{fig:vrot_rextent_sn_check}%
	The fractional deviation of the $V_{\rmn{rot},j}$ of individual
	lines, from the weighted mean of all the `good' lines for that 
	galaxy ($\vrot$), plotted versus (a) $\rextent$ and (b) \snr.
	Filled points correspond to emission lines from cluster galaxies,
	while open points are from field galaxies.
	Points on the dotted line have a fractional deviation of zero,
	usually because only one emission line is available for that galaxy.
	Note that while the scatter varies, the plots indicate no correlation
	between the rotation velocity measured from a line and that line's
	$\rextent$ or \snr.
}
\end{figure}

In addition, even if there are differences in the $\rextent$ and
\snr{} distributions of our cluster and field samples, we find no
correlation between $\vrot$ and these quantities, so this cannot be
responsible for the TFR offset we measure.  This is demonstrated by
Fig.~\ref{fig:vrot_rextent_sn_check}, plots of the fractional
deviation of the $V_{\rmn{rot},j}$ of individual lines, from the
weighted mean of all the `good' lines for that galaxy ($\vrot$),
versus $\rextent$ and \snr.  Note that the scatter increases with both
decreasing $\rextent$ and \snr, as one would expect.

\begin{figure*}%
\centering%
\includegraphics[width=\textwidth]%
		{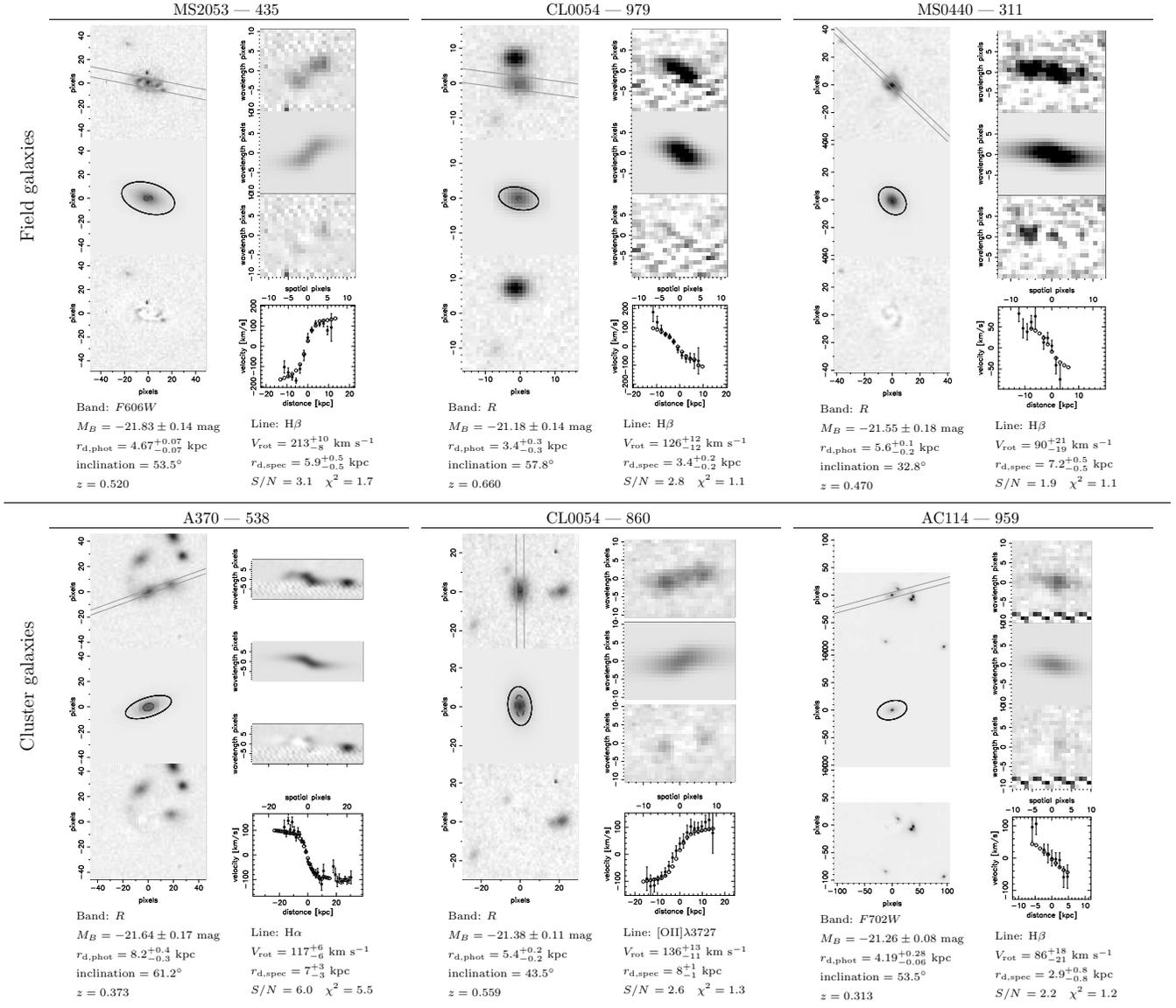}%
\caption{\label{fig:examples}%
	Representative examples of our data, models and 
	observed rotation curves.
	Examples are shown for three galaxies from each of the matched TFR
	field (top row) and cluster (bottom row) samples.
	These example galaxies have been selected by their $\vrot$ errors,
	such that on each row, from left to right, we show the galaxy with the 
	10th-, 50th- and 90th-percentile $\sigma_{\vrot}$.
	For each galaxy, labelled above by cluster name and id number,
	two columns are given.  Their contents, from top to bottom, are:
	\emph{Left --- example imaging data:}
	\emph{(top)} the best HST or $R$-band image available
	with the spectroscopic slit overlaid,
	\emph{(middle)} \gimd{} model (greyscale) of the above image with
	ellipses overlaid indicating $3\rdphot$ (black line)
	and the bulge effective radius (grey line, dashed if bulge fraction $< 0.1$),
	\emph{(bottom)} data$-$model residual image,
	\emph{(text)} band of the shown image, absolute rest-frame $B$-band magnitude,
	photometric disc scalelength, adopted inclination and redshift.
	\emph{Right --- example spectroscopic data:}
	\emph{(top)} the emission line contributing most to the $\vrot$ measurement,
	\emph{(middle)} \elfitpy{} model of the above emission line,
	\emph{(bottom)} data$-$model residual image,
	\emph{(plot)} the observed rotation curve (filled points) and model
	rotation curve (as observed, open points), including the effects of inclination,
	seeing, etc., from a combination of all the available emission lines,
	\emph{(text)} wavelength of the example line, (rest-frame) rotation velocity and
	spectroscopic emission scalelength (possibly combined from fits to multiple 
	lines), \snr and $\chi_{\rmn{r}}^2$ for the example line.}
\end{figure*}

Representative examples of our data, model lines and observed rotation
curves, for both field and cluster galaxies, are shown in
Fig.~\ref{fig:examples}.  The plotted rotation curves have been
measured by the tracing method described in \S\ref{sec:spec}, combined
by weighted averages of the reliable points in the case of multiple
lines for a single galaxy.  Only points with at least one `good'
measurement are plotted, thereby showing the extent to which we can
reliably trace the line.  The model lines have been traced, and `good'
points determined, in exactly the same way, so that the extent of the
model line shows the distance to which it can be reliably traced
assuming the same pixel errors as the real data.
Note that $\vrot$ is not measured using this method, but rather by
comparison with model 2D spectra in the Metropolis parameter search of
\elfitpy.  Visually there is no difference in the form and quality of
the emission lines and rotation curves between the two samples.  We
therefore assert that the offset between the TFR of the two samples is
due to real, physical differences in $M_B$ and/or $\vrot$.

Note that the objects in our sample are giant galaxies, which must
have emission lines bright enough for us to be able to fit and hence
measure $\vrot$.  Our results therefore apply to massive ($\vrot \ga
80$ \kms), luminous ($M_B > -19.5$) galaxies, with significant active
star-formation in the disc.  No conclusions may be drawn concerning
the population of fainter disc galaxies or those with little or no
ongoing star-formation.

For the MS1054 sample alone, evidence for a correlation between $\dtf$
and star-formation rate has already been demonstrated
(\citealt{MJthesis}, Milvang-Jensen \& Arag{\' o}n-Salamanca in
preparation). This suggests that a change in $M_B$ due to enhanced
star-formation is the main cause of the TFR cluster$-$field offset.
More insight into this issue will be provided by future investigation
of the star-formation rates for the galaxies in the present sample,
combined with those from our Subaru study.

\vspace{-0.5\baselineskip}
\subsection{Comparison with other studies}

In contrast to our result, the study of \citet{ZBJHM03} finds no
difference between the TFR of $13$ spirals in three clusters at $0.3
\la z \la 0.5$ and that determined for the FORS Deep Field by
\citet{Bohmetal04}.  This is puzzling, and may point to real
differences in TFR offsets between individual clusters. However, this
question must wait to be addressed by larger studies which can examine
TFR offsets, along with SFRs and colours, as a function of cluster
properties.

It seems difficult to attribute the conflict between our results and
those of \citet{ZBJHM03} to a difference in sample selection
\citep[see][]{Jager04}.  This was performed in a fairly similar
manner, generally giving preference to galaxies based on luminosity,
spiral morphology, known emission lines, and cluster membership.
However, both studies have rather heterogeneous selection procedures,
based upon the availability of disparate prior data in the literature.
Additional, higher quality data for four more clusters are expected
soon from this group, which should help confirm which result is
correct.  We are also in the process of fitting subsamples of each
others' emission line data, in order to assess the robustness of our
rotation velocities with respect to the analysis methods used.

A study of $15$ spirals in the cluster CL0024+1654 at $z=0.4$ by
\citet{MK04} finds a TFR offset for their galaxies, with respect to
the \emph{local} field, of $\unisim 0.5$ mag.  This suggests there is
little difference between cluster and field spirals at $z \sim 0.4$,
when combined with the $\unisim 0.4$ mag of field evolution expected
at this redshift from the results of \papertwo.  However, few details
of their study have been published to date, and so a proper comparison
with our work is not possible.  Data for more clusters will be
provided soon by our Subaru study (Nakamura et al. in preparation),
and future work by the EDisCS collaboration.

More general studies of the correlation between SFR and local galaxy
density by \citet{Lewis02} and \citet{Gomez03}, using the 2dF and SDSS
datasets respectively, both find the existence of a critical local
galaxy density (of $\unisim 1$ galaxy with $M_b \la -19$ per Mpc$^2$).
At densities greater than this, the average SFR decreases with
density.  At lower densities there is no significant correlation.
These results imply that the global SFR of the universe may be
influenced by environmental effects at quite low densities, outside of
the boundaries of rich clusters, and therefore its variation is not
simply due to a general, internal evolution of the SFR in individual
galaxies.

How does the finding that groups may be the dominant site of
star-formation suppression today compare with our result, that we also
find this process occurring -- accompanied by a SFR \emph{enhancement}
-- in rich clusters at intermediate redshift?  Firstly, the fact that
suppression of SFR happens at low densities locally does not rule out
it also occurring in cluster environments.  Rather, the linearity of
the SFR--density correlation implies that the efficiency of SFR
suppression increases with density. 

Furthermore, \citet{Betal04} find that the environmental dependence of
the volume averaged SFR is due to changing proportions of the
star-forming and passive galaxy populations, rather than a shift in
the mean SFR of star-forming galaxies.  This suggests that the process
responsible for reducing the average SFR in groups is
\emph{stochastic}.  When the process occurs it causes a halt in
star-formation, and hence a transformation from a galaxy in the
star-forming population into the passive population.  However, in
order to preserve the smooth correlation between SFR and local
density, this process must occur randomly, with a frequency related to
the local density.  This suggests mergers as the responsible mechanism
for SFR suppression in groups.  However, in local rich clusters very
few star-forming galaxies are found, yet mergers are less likely due
to the large relative galaxy velocities.  In this case it may be that
a more all-encompassing mechanism, such as ram-pressure stripping, is
at work, finishing the job started in groups.

Another finding by \citet{Betal04}, that star-forming galaxies in
dense environments have an EW(H$\alpha$) distribution
indistinguishable from that for low-density environments, appears at
first to be inconsistent with the present paper's results.  However,
the necessarily short time-scale for any SFR enhancement, combined
with the simultaneous existence of galaxies with declining SFR, may
make detecting such an effect difficult using the EW(H$\alpha$)
distribution.

A further explanation may be one of pre-processing.  It seems likely
that galaxies falling into rich clusters today have spent a longer
time subjected to group conditions than those entering similar
clusters at $z\sim0.5$.  If, as is suggested above, the probability
for star-formation suppression increases with both local density and
the length of time which the galaxy has been subjected to the
environment, we would therefore expect clusters to be the site of
star-formation truncation at intermediate redshifts, but no longer
today -- at least for massive galaxies, which are preferentially
located in denser regions.  However, to assert this will require a
consideration of cosmological simulations beyond the scope of the
present paper.

There has been surprisingly little direct study of the local ($z\sim0$)
dependence of the TFR on environment, although this is perhaps because
a lack of any dependence is apparent in more general studies.  An
investigation by \citet{BGMM90} finds no evidence for a difference
between the TFR of spirals in clusters and those in a sample taken
from groups and the field.  This provides some evidence that any
difference between cluster and field spirals that may have existed in
the past, has now diminished, at least for those spirals which retain
significant quantities of HI.  Studies of asymmetry, truncation and HI
deficiency in cluster spirals have also been performed for local
clusters \citep{Dale01,Vogt04II}, finding evidence for the stripping
of disc gas through some process related to galaxy infall.

The variation of galaxy properties with environment, as investigated
by the studies mentioned above, suggests a similar examination of TFR
offset with respect to clustercentric distance and local density for
our data.  We plan to undertake a such a detailed `geographical' study
of our VLT and Subaru intermediate-redshift clusters once the samples
have been combined and further spectral properties measured.  We
therefore do not consider this any further in the present paper.

\vspace{-1.2\baselineskip}
\section{Conclusions}

We have constructed the TFR for samples of homogeneously selected and
analysed field and cluster spirals with $0.25 \leq z \leq 1$ and $M_B
\leq -19.5$.  From this we have found that cluster galaxies are offset
from the field TFR by $0.7\pm0.2$ mag, such that those in clusters are
over-luminous for a given rotation velocity.  The reality of this
offset is significant at a $3\sigma$ confidence level.  This offset
remains, with similar significance, even if a global evolution in the
field population is taken into account.
It should be stressed that this result applies only to the bright,
massive, star-forming, disc galaxies which form the sample considered
here.  However, we do find a marginal indication that the galaxies in
our sample with lower rotation velocities ($\la 150$ \kms) contribute
most to our measured offset.

We have extensively compared the emission lines of the field and
cluster samples, finding no difference to which we could attribute the
TFR offset.  The most likely explanation is that the cluster galaxies
have been brightened by their initial interaction with the
intra-cluster medium.  This is presumably due to an initial
enhancement of their star-formation rate, before further interaction
has suppressed it.
However, at this point we cannot rule out the possibility of a change
in $\vrot$ due to stripping of the dark matter haloes of cluster
galaxies.  Discriminating between TFR offsets due to changes in either
luminosity or rotation velocity will require further work, such as
examining the difference in star-formation rate for distant cluster
and field galaxies, and its correlation with TFR offset.

\vspace{-1.2\baselineskip}
\section*{Acknowledgments}
We would like to thank Ian Smail, for generously providing additional
imaging data, and Pierre-Alain Duc, for allowing us access to an
unpublished spectroscopic catalogue of CL0054 used in the target
selection process.
We also thank Osamu Nakamura for useful discussions, and the referee,
Bodo L. Ziegler, for his detailed and helpful comments.
The Simbad and VizieR services provided by the Centre de Donn\'ees
astronomiques de Strasbourg (CDS) were used in the course of this work.

\vspace{-0.75\baselineskip}
\bsp

\vspace{-1.2\baselineskip}
{\small%
\setlength{\baselineskip}{0.90\baselineskip}

}

\appendix
\section{Additional photometry information}

\begin{table*}%
\begin{minipage}{0.86\textwidth}%
\caption{\label{tab:phot_info}%
	Details of the imaging availability for each galaxy in our
	full TFR sample.  The symbols in columns headed by a
	photometric band designation indicate the bands in which we
	were able to measure magnitudes. Numerical band designations
	correspond to HST/WFPC2 filters, e.g. $F555W$.  The symbol
	\bandyes{} indicates that magnitude information in this band
	was available, and \bandrf{} additionally specifies the band
	which formed the basis for the conversion to rest-frame
	$B$-band magnitude.  The columns headed `$i$' and
	`$r_{\rmn{d}}$' indicate whether the inclinations and
	photometric scalelengths are based on HST (\inchst) or
	ground-based (\incground) imaging.}%
\centering%
\begin{tabular}{l@{}c@{}c@{}c@{}c@{}c@{}c@{}c@{}c@{}c@{}c@{}c@{}c@{}c@{}c@{}c@{}}
\hline
&&&\multicolumn{9}{c}{\makebox[8.5\photinfocolwidth]{\hrulefill Bands with magnitude available\hrulefill}}&
&\multicolumn{2}{c}{\hrulefill HST\hrulefill}\\
\makebox[2.5\photinfocolwidth]{ID} & 
\makebox[1.5\photinfocolwidth]{$z$} & \makebox[1.5\photinfocolwidth]{Mem.} &
\makebox[\photinfocolwidth]{$B$} &
\makebox[\photinfocolwidth]{$555$} & \makebox[\photinfocolwidth]{$V$} &
\makebox[\photinfocolwidth]{$606$} &
\makebox[\photinfocolwidth]{$R$} & \makebox[\photinfocolwidth]{$675$} &
\makebox[\photinfocolwidth]{$702$} & \makebox[\photinfocolwidth]{$I$} &
\makebox[\photinfocolwidth]{$814$} & \makebox[0.5\photinfocolwidth]{} &
\makebox[\photinfocolwidth]{$i$} & \makebox[\photinfocolwidth]{$r_{\rmn{d}}$}\\
\hline
      MS0440\_101 & $   0.819$ &     \memno &    \bandno &    \bandno &    \bandno &    \bandno &    \bandrf &    \bandno &    \bandno &    \bandno &    \bandno && \incground &  \rdground\\
      MS0440\_140 & $   0.316$ &     \memno &    \bandno &    \bandno &    \bandno &    \bandno &    \bandrf &    \bandno &    \bandno &    \bandno &    \bandno && \incground &  \rdground\\
      MS0440\_188 & $   0.491$ &     \memno &    \bandno &    \bandno &    \bandno &    \bandno &    \bandrf &    \bandno &    \bandno &    \bandno &    \bandno && \incground &  \rdground\\
      MS0440\_207 & $   0.087$ &     \memno &    \bandno &    \bandno &    \bandno &    \bandno &    \bandrf &    \bandno &    \bandno &    \bandno &    \bandno && \incground &  \rdground\\
      MS0440\_273 & $   0.283$ &     \memno &    \bandno &    \bandno &    \bandno &    \bandno &    \bandrf &    \bandno &    \bandno &    \bandno &    \bandno && \incground &  \rdground\\
      MS0440\_311 & $   0.470$ &     \memno &    \bandno &    \bandno &    \bandno &    \bandno &    \bandrf &    \bandno &    \bandno &   \bandyes &    \bandno && \incground &  \rdground\\
      MS0440\_319 & $   0.138$ &     \memno &    \bandno &    \bandno &    \bandno &    \bandno &    \bandrf &    \bandno &    \bandno &   \bandyes &    \bandno && \incground &  \rdground\\
      MS0440\_538 & $   0.213$ &     \memno &    \bandno &    \bandno &    \bandno &    \bandno &    \bandrf &    \bandno &    \bandno &   \bandyes &    \bandno && \incground &  \rdground\\
      MS0440\_616 & $   0.211$ &     \memno &    \bandno &    \bandno &    \bandno &    \bandno &    \bandrf &    \bandno &    \bandno &   \bandyes &    \bandno && \incground &  \rdground\\
      MS0440\_627 & $   0.265$ &     \memno &    \bandno &    \bandno &    \bandno &    \bandno &    \bandrf &    \bandno &    \bandno &   \bandyes &    \bandno && \incground &  \rdground\\
      MS0440\_635 & $   0.237$ &     \memno &    \bandno &    \bandno &    \bandno &    \bandno &    \bandrf &    \bandno &    \bandno &    \bandno &    \bandno && \incground &  \rdground\\
      MS0440\_657 & $   0.265$ &     \memno &    \bandno &    \bandno &    \bandno &    \bandno &    \bandrf &    \bandno &    \bandno &    \bandno &    \bandno && \incground &  \rdground\\
      MS0440\_735 & $   0.181$ &     \memno &    \bandno &    \bandno &    \bandno &    \bandno &    \bandrf &    \bandno &   \bandyes &   \bandyes &    \bandno &&    \inchst &     \rdhst\\
      MS0440\_849 & $   0.401$ &     \memno &    \bandno &    \bandno &    \bandno &    \bandno &    \bandrf &    \bandno &    \bandno &   \bandyes &    \bandno && \incground &  \rdground\\
     MS0440\_1109 & $   0.239$ &     \memno &    \bandno &    \bandno &    \bandno &    \bandno &    \bandrf &    \bandno &   \bandyes &   \bandyes &    \bandno &&    \inchst &     \rdhst\\
     MS0440\_1131 & $   0.318$ &     \memno &    \bandno &    \bandno &    \bandno &    \bandno &    \bandrf &    \bandno &    \bandno &   \bandyes &    \bandno && \incground &  \rdground\\
     MS0440\_1157 & $   0.401$ &     \memno &    \bandno &    \bandno &    \bandno &    \bandno &    \bandrf &    \bandno &    \bandno &   \bandyes &    \bandno && \incground &  \rdground\\
       AC114\_115 & $   0.500$ &     \memno &   \bandyes &    \bandno &    \bandno &    \bandno &    \bandrf &    \bandno &    \bandno &   \bandyes &    \bandno && \incground &  \rdground\\
       AC114\_264 & $   0.098$ &     \memno &    \bandrf &    \bandno &    \bandno &    \bandno &   \bandyes &    \bandno &   \bandyes &   \bandyes &    \bandno &&    \inchst &     \rdhst\\
       AC114\_391 & $   0.567$ &     \memno &   \bandyes &    \bandno &    \bandno &    \bandno &   \bandyes &    \bandno &    \bandrf &   \bandyes &    \bandno &&    \inchst &     \rdhst\\
       AC114\_553 & $   0.210$ &     \memno &    \bandno &    \bandno &    \bandno &    \bandno &    \bandrf &    \bandno &   \bandyes &    \bandno &    \bandno && \incground &  \rdground\\
       AC114\_700 & $   0.351$ &     \memno &   \bandyes &    \bandno &    \bandno &    \bandno &    \bandrf &    \bandno &   \bandyes &   \bandyes &    \bandno &&    \inchst &     \rdhst\\
       AC114\_810 & $   0.354$ &     \memno &   \bandyes &    \bandno &    \bandno &    \bandno &    \bandrf &    \bandno &   \bandyes &   \bandyes &    \bandno &&    \inchst &     \rdhst\\
       AC114\_875 & $   0.171$ &     \memno &    \bandrf &    \bandno &    \bandno &    \bandno &   \bandyes &    \bandno &   \bandyes &   \bandyes &    \bandno &&    \inchst &     \rdhst\\
         A370\_39 & $   0.325$ &     \memno &    \bandno &    \bandno &    \bandno &    \bandno &    \bandrf &    \bandno &    \bandno &    \bandno &    \bandno && \incground &  \rdground\\
        A370\_119 & $   0.564$ &     \memno &    \bandno &   \bandyes &    \bandno &    \bandno &    \bandrf &    \bandno &    \bandno &   \bandyes &   \bandyes &&    \inchst &  \rdground\\
        A370\_157 & $   0.542$ &     \memno &    \bandno &    \bandno &    \bandno &    \bandno &    \bandrf &    \bandno &    \bandno &   \bandyes &    \bandno && \incground &  \rdground\\
        A370\_183 & $   0.361$ &     \memno &    \bandno &   \bandyes &    \bandno &    \bandno &    \bandrf &    \bandno &    \bandno &   \bandyes &   \bandyes &&    \inchst &  \rdground\\
        A370\_210 & $   0.230$ &     \memno &    \bandno &    \bandrf &    \bandno &    \bandno &   \bandyes &    \bandno &    \bandno &   \bandyes &   \bandyes &&    \inchst &  \rdground\\
        A370\_292 & $   0.542$ &     \memno &    \bandno &   \bandyes &    \bandno &    \bandno &    \bandno &    \bandno &    \bandno &    \bandrf &   \bandyes &&    \inchst &  \rdground\\
        A370\_319 & $   0.305$ &     \memno &    \bandno &    \bandno &    \bandno &    \bandno &    \bandrf &   \bandyes &    \bandno &   \bandyes &    \bandno &&    \inchst &     \rdhst\\
        A370\_401 & $   0.346$ &     \memno &    \bandno &    \bandno &    \bandno &    \bandno &    \bandrf &   \bandyes &    \bandno &   \bandyes &    \bandno &&    \inchst &     \rdhst\\
        A370\_406 & $   0.571$ &     \memno &    \bandno &    \bandno &    \bandno &    \bandno &    \bandrf &    \bandno &    \bandno &    \bandno &    \bandno && \incground &  \rdground\\
        A370\_540 & $   0.173$ &     \memno &    \bandno &    \bandno &    \bandno &    \bandno &    \bandrf &    \bandno &    \bandno &    \bandno &    \bandno && \incground &  \rdground\\
        A370\_582 & $   0.207$ &     \memno &    \bandno &    \bandno &    \bandno &    \bandno &    \bandrf &    \bandno &    \bandno &    \bandno &    \bandno && \incground &  \rdground\\
        A370\_620 & $   0.250$ &     \memno &    \bandno &    \bandno &    \bandno &    \bandno &    \bandrf &    \bandno &    \bandno &    \bandno &    \bandno && \incground &  \rdground\\
        A370\_630 & $   0.225$ &     \memno &    \bandno &    \bandno &    \bandno &    \bandno &    \bandrf &   \bandyes &    \bandno &   \bandyes &    \bandno &&    \inchst &     \rdhst\\
        A370\_650 & $   0.547$ &     \memno &    \bandno &    \bandno &    \bandno &    \bandno &    \bandrf &    \bandno &    \bandno &   \bandyes &    \bandno && \incground &  \rdground\\
        A370\_751 & $   0.256$ &     \memno &    \bandno &    \bandno &    \bandno &    \bandno &    \bandrf &   \bandyes &    \bandno &   \bandyes &    \bandno &&    \inchst &     \rdhst\\
       CL0054\_62 & $   0.537$ &     \memno &    \bandno &    \bandno &    \bandno &    \bandno &    \bandrf &    \bandno &    \bandno &    \bandno &    \bandno && \incground &  \rdground\\
       CL0054\_83 & $   0.718$ &     \memno &    \bandno &    \bandno &    \bandno &    \bandno &    \bandrf &    \bandno &    \bandno &    \bandno &    \bandno && \incground &  \rdground\\
       CL0054\_89 & $   0.537$ &     \memno &    \bandno &    \bandno &    \bandno &    \bandno &    \bandrf &    \bandno &    \bandno &    \bandno &    \bandno && \incground &  \rdground\\
      CL0054\_126 & $   0.237$ &     \memno &    \bandno &    \bandno &    \bandno &    \bandno &    \bandrf &    \bandno &    \bandno &    \bandno &    \bandno && \incground &  \rdground\\
      CL0054\_137 & $   0.297$ &     \memno &    \bandno &    \bandno &    \bandno &    \bandno &    \bandrf &    \bandno &    \bandno &    \bandno &    \bandno && \incground &  \rdground\\
      CL0054\_138 & $   0.237$ &     \memno &    \bandno &    \bandno &    \bandno &    \bandno &    \bandrf &    \bandno &    \bandno &    \bandno &    \bandno && \incground &  \rdground\\
      CL0054\_284 & $   0.815$ &     \memno &    \bandno &    \bandno &    \bandno &    \bandno &    \bandrf &    \bandno &    \bandno &    \bandno &    \bandno && \incground &  \rdground\\
      CL0054\_354 & $   0.224$ &     \memno &    \bandno &    \bandno &    \bandrf &    \bandno &   \bandyes &    \bandno &    \bandno &   \bandyes &    \bandno && \incground &  \rdground\\
\hline
\end{tabular}
\end{minipage}
\end{table*}


\begin{table*}%
\begin{minipage}{0.86\textwidth}%
\contcaption{\label{lastpage}}%
\centering%
\begin{tabular}{l@{}c@{}c@{}c@{}c@{}c@{}c@{}c@{}c@{}c@{}c@{}c@{}c@{}c@{}c@{}c@{}}
\hline
&&&\multicolumn{9}{c}{\makebox[8.5\photinfocolwidth]{\hrulefill Bands with magnitude available\hrulefill}}&
&\multicolumn{2}{c}{\hrulefill HST\hrulefill}\\
\makebox[2.5\photinfocolwidth]{ID} & 
\makebox[1.5\photinfocolwidth]{$z$} & \makebox[1.5\photinfocolwidth]{Mem.} &
\makebox[\photinfocolwidth]{$B$} &
\makebox[\photinfocolwidth]{$555$} & \makebox[\photinfocolwidth]{$V$} &
\makebox[\photinfocolwidth]{$606$} &
\makebox[\photinfocolwidth]{$R$} & \makebox[\photinfocolwidth]{$675$} &
\makebox[\photinfocolwidth]{$702$} & \makebox[\photinfocolwidth]{$I$} &
\makebox[\photinfocolwidth]{$814$} & \makebox[0.5\photinfocolwidth]{} &
\makebox[\photinfocolwidth]{$i$} & \makebox[\photinfocolwidth]{$r_{\rmn{d}}$}\\
\hline
      CL0054\_407 & $   0.275$ &     \memno &    \bandno &   \bandyes &    \bandrf &    \bandno &   \bandyes &    \bandno &    \bandno &   \bandyes &   \bandyes &&    \inchst &  \rdground\\
      CL0054\_454 & $   0.298$ &     \memno &    \bandno &    \bandno &    \bandno &    \bandno &    \bandrf &    \bandno &    \bandno &    \bandno &    \bandno && \incground &  \rdground\\
      CL0054\_579 & $   0.577$ &     \memno &    \bandno &    \bandno &    \bandno &    \bandno &    \bandrf &    \bandno &    \bandno &    \bandno &    \bandno && \incground &  \rdground\\
      CL0054\_588 & $   0.911$ &     \memno &    \bandno &    \bandno &    \bandno &    \bandno &    \bandrf &    \bandno &    \bandno &    \bandno &    \bandno && \incground &  \rdground\\
      CL0054\_686 & $   0.710$ &     \memno &    \bandno &    \bandno &    \bandno &    \bandno &    \bandrf &    \bandno &    \bandno &    \bandno &    \bandno && \incground &  \rdground\\
      CL0054\_688 & $   0.298$ &     \memno &    \bandno &    \bandno &    \bandno &    \bandno &    \bandrf &    \bandno &    \bandno &    \bandno &    \bandno && \incground &  \rdground\\
      CL0054\_779 & $   1.004$ &     \memno &    \bandno &    \bandno &    \bandno &    \bandno &    \bandrf &    \bandno &    \bandno &    \bandno &    \bandno && \incground &  \rdground\\
      CL0054\_803 & $   0.162$ &     \memno &    \bandno &    \bandno &    \bandno &    \bandno &    \bandrf &    \bandno &    \bandno &    \bandno &    \bandno && \incground &  \rdground\\
      CL0054\_827 & $   0.583$ &     \memno &    \bandno &    \bandno &    \bandno &    \bandno &    \bandrf &    \bandno &    \bandno &    \bandno &    \bandno && \incground &  \rdground\\
      CL0054\_892 & $   0.585$ &     \memno &    \bandno &    \bandno &   \bandyes &    \bandno &    \bandrf &    \bandno &    \bandno &   \bandyes &    \bandno && \incground &  \rdground\\
      CL0054\_927 & $   0.653$ &     \memno &    \bandno &    \bandno &    \bandno &    \bandno &    \bandrf &    \bandno &    \bandno &    \bandno &    \bandno && \incground &  \rdground\\
      CL0054\_937 & $   0.603$ &     \memno &    \bandno &   \bandyes &   \bandyes &    \bandno &    \bandrf &    \bandno &    \bandno &   \bandyes &   \bandyes &&    \inchst &  \rdground\\
      CL0054\_979 & $   0.660$ &     \memno &    \bandno &    \bandno &    \bandno &    \bandno &    \bandrf &    \bandno &    \bandno &    \bandno &    \bandno && \incground &  \rdground\\
      CL0054\_993 & $   0.214$ &     \memno &    \bandno &    \bandno &    \bandno &    \bandno &    \bandrf &    \bandno &    \bandno &    \bandno &    \bandno && \incground &  \rdground\\
     CL0054\_1011 & $   0.171$ &     \memno &    \bandno &    \bandrf &   \bandyes &    \bandno &   \bandyes &    \bandno &    \bandno &   \bandyes &   \bandyes &&    \inchst &  \rdground\\
     CL0054\_1054 & $   0.830$ &     \memno &    \bandno &    \bandno &    \bandno &    \bandno &    \bandrf &    \bandno &    \bandno &    \bandno &    \bandno && \incground &  \rdground\\
       MS2053\_86 & $   0.196$ &     \memno &    \bandno &    \bandno &    \bandno &    \bandno &    \bandrf &    \bandno &    \bandno &    \bandno &    \bandno && \incground &  \rdground\\
      MS2053\_371 & $   0.521$ &     \memno &    \bandno &    \bandno &    \bandno &   \bandyes &    \bandrf &    \bandno &    \bandno &    \bandno &   \bandyes &&    \inchst &     \rdhst\\
      MS2053\_404 & $   0.384$ &     \memno &    \bandno &    \bandno &    \bandno &    \bandrf &   \bandyes &    \bandno &    \bandno &    \bandno &   \bandyes &&    \inchst &     \rdhst\\
      MS2053\_435 & $   0.520$ &     \memno &    \bandno &    \bandno &    \bandno &    \bandrf &    \bandno &    \bandno &    \bandno &    \bandno &   \bandyes &&    \inchst &     \rdhst\\
      MS2053\_455 & $   0.174$ &     \memno &    \bandno &    \bandno &    \bandno &    \bandrf &   \bandyes &    \bandno &    \bandno &    \bandno &   \bandyes &&    \inchst &     \rdhst\\
      MS2053\_470 & $   0.371$ &     \memno &    \bandno &    \bandno &    \bandno &    \bandrf &   \bandyes &    \bandno &    \bandno &    \bandno &   \bandyes &&    \inchst &     \rdhst\\
      MS2053\_741 & $   0.335$ &     \memno &    \bandno &    \bandno &    \bandno &    \bandrf &   \bandyes &    \bandno &    \bandno &    \bandno &    \bandno &&    \inchst &     \rdhst\\
      MS2053\_856 & $   0.261$ &     \memno &    \bandno &    \bandno &    \bandno &    \bandrf &   \bandyes &    \bandno &    \bandno &    \bandno &    \bandno &&    \inchst &     \rdhst\\
      MS2053\_998 & $   0.196$ &     \memno &    \bandno &    \bandno &    \bandno &    \bandrf &   \bandyes &    \bandno &    \bandno &    \bandno &   \bandyes &&    \inchst &     \rdhst\\
     MS2053\_1105 & $   0.408$ &     \memno &    \bandno &    \bandno &    \bandno &    \bandrf &   \bandyes &    \bandno &    \bandno &    \bandno &   \bandyes &&    \inchst &     \rdhst\\
     MS2053\_1296 & $   0.058$ &     \memno &    \bandno &    \bandno &    \bandno &    \bandno &    \bandrf &    \bandno &    \bandno &    \bandno &    \bandno && \incground &  \rdground\\
      MS1054\_F02 & $   0.180$ &     \memno &    \bandno &    \bandno &    \bandno &    \bandrf &    \bandno &    \bandno &    \bandno &    \bandno &   \bandyes &&    \inchst &     \rdhst\\
      MS1054\_F04 & $   0.230$ &     \memno &    \bandno &    \bandno &    \bandno &    \bandrf &    \bandno &    \bandno &    \bandno &    \bandno &   \bandyes &&    \inchst &     \rdhst\\
      MS1054\_F05 & $   0.249$ &     \memno &    \bandno &    \bandno &    \bandno &    \bandrf &    \bandno &    \bandno &    \bandno &    \bandno &   \bandyes &&    \inchst &     \rdhst\\
      MS1054\_F06 & $   0.259$ &     \memno &    \bandno &    \bandno &    \bandno &    \bandrf &    \bandno &    \bandno &    \bandno &    \bandno &   \bandyes &&    \inchst &     \rdhst\\
      MS1054\_F08 & $   0.287$ &     \memno &    \bandno &    \bandno &    \bandno &    \bandrf &    \bandno &    \bandno &    \bandno &    \bandno &   \bandyes &&    \inchst &     \rdhst\\
      MS1054\_F10 & $   0.324$ &     \memno &    \bandno &    \bandno &    \bandno &    \bandrf &    \bandno &    \bandno &    \bandno &    \bandno &   \bandyes &&    \inchst &     \rdhst\\
      MS1054\_F11 & $   0.325$ &     \memno &    \bandno &    \bandno &    \bandno &    \bandrf &    \bandno &    \bandno &    \bandno &    \bandno &   \bandyes &&    \inchst &     \rdhst\\
      MS1054\_F12 & $   0.325$ &     \memno &    \bandno &    \bandno &    \bandno &    \bandrf &    \bandno &    \bandno &    \bandno &    \bandno &   \bandyes &&    \inchst &     \rdhst\\
      MS1054\_F14 & $   0.429$ &     \memno &    \bandno &    \bandno &    \bandno &    \bandrf &    \bandno &    \bandno &    \bandno &    \bandno &    \bandno &&    \inchst &     \rdhst\\
      MS1054\_F16 & $   0.470$ &     \memno &    \bandno &    \bandno &    \bandno &    \bandrf &    \bandno &    \bandno &    \bandno &    \bandno &   \bandyes &&    \inchst &     \rdhst\\
      MS1054\_F18 & $   0.553$ &     \memno &    \bandno &    \bandno &    \bandno &    \bandrf &    \bandno &    \bandno &    \bandno &    \bandno &   \bandyes &&    \inchst &     \rdhst\\
      MS1054\_F19 & $   0.684$ &     \memno &    \bandno &    \bandno &    \bandno &   \bandyes &    \bandno &    \bandno &    \bandno &    \bandno &    \bandrf &&    \inchst &     \rdhst\\
      MS1054\_F20 & $   0.686$ &     \memno &    \bandno &    \bandno &    \bandno &   \bandyes &    \bandno &    \bandno &    \bandno &    \bandno &    \bandrf &&    \inchst &     \rdhst\\
      MS1054\_F21 & $   0.756$ &     \memno &    \bandno &    \bandno &    \bandno &   \bandyes &    \bandno &    \bandno &    \bandno &    \bandno &    \bandrf &&    \inchst &     \rdhst\\
      MS1054\_F22 & $   0.896$ &     \memno &    \bandno &    \bandno &    \bandno &   \bandyes &    \bandno &    \bandno &    \bandno &    \bandno &    \bandrf &&    \inchst &     \rdhst\\
        AC114\_18 & $   0.306$ &    \memyes &   \bandyes &    \bandno &    \bandno &    \bandno &    \bandrf &    \bandno &    \bandno &   \bandyes &    \bandno && \incground &  \rdground\\
       AC114\_142 & $   0.325$ &    \memyes &   \bandyes &    \bandno &    \bandno &    \bandno &    \bandrf &    \bandno &    \bandno &   \bandyes &    \bandno && \incground &  \rdground\\
       AC114\_193 & $   0.307$ &    \memyes &   \bandyes &    \bandno &    \bandno &    \bandno &    \bandrf &    \bandno &   \bandyes &   \bandyes &    \bandno &&    \inchst &     \rdhst\\
       AC114\_768 & $   0.314$ &    \memyes &   \bandyes &    \bandno &    \bandno &    \bandno &    \bandrf &    \bandno &    \bandno &   \bandyes &    \bandno && \incground &  \rdground\\
       AC114\_930 & $   0.306$ &    \memyes &   \bandyes &    \bandno &    \bandno &    \bandno &    \bandrf &    \bandno &   \bandyes &   \bandyes &    \bandno &&    \inchst &     \rdhst\\
       AC114\_959 & $   0.313$ &    \memyes &   \bandyes &    \bandno &    \bandno &    \bandno &    \bandrf &    \bandno &   \bandyes &   \bandyes &    \bandno &&    \inchst &     \rdhst\\
      AC114\_1001 & $   0.307$ &    \memyes &   \bandyes &    \bandno &    \bandno &    \bandno &    \bandrf &    \bandno &    \bandno &   \bandyes &    \bandno && \incground &  \rdground\\
        A370\_532 & $   0.374$ &    \memyes &    \bandno &    \bandno &    \bandno &    \bandno &    \bandrf &    \bandno &    \bandno &   \bandyes &    \bandno && \incground &  \rdground\\
        A370\_538 & $   0.373$ &    \memyes &    \bandno &    \bandno &    \bandno &    \bandno &    \bandrf &    \bandno &    \bandno &   \bandyes &    \bandno && \incground &  \rdground\\
        A370\_555 & $   0.378$ &    \memyes &    \bandno &    \bandno &    \bandno &    \bandno &    \bandrf &    \bandno &    \bandno &    \bandno &    \bandno && \incground &  \rdground\\
      CL0054\_358 & $   0.564$ &    \memyes &    \bandno &    \bandno &   \bandyes &    \bandno &    \bandrf &    \bandno &    \bandno &   \bandyes &    \bandno && \incground &  \rdground\\
      CL0054\_609 & $   0.558$ &    \memyes &    \bandno &    \bandno &    \bandno &    \bandno &    \bandrf &    \bandno &    \bandno &    \bandno &    \bandno && \incground &  \rdground\\
      CL0054\_643 & $   0.558$ &    \memyes &    \bandno &    \bandno &    \bandno &    \bandno &    \bandrf &    \bandno &    \bandno &    \bandno &    \bandno && \incground &  \rdground\\
      CL0054\_714 & $   0.562$ &    \memyes &    \bandno &    \bandno &    \bandno &    \bandno &    \bandrf &    \bandno &    \bandno &    \bandno &    \bandno && \incground &  \rdground\\
      CL0054\_725 & $   0.557$ &    \memyes &    \bandno &    \bandno &    \bandno &    \bandno &    \bandrf &    \bandno &    \bandno &    \bandno &    \bandno && \incground &  \rdground\\
      CL0054\_799 & $   0.554$ &    \memyes &    \bandno &    \bandno &    \bandno &    \bandno &    \bandrf &    \bandno &    \bandno &    \bandno &    \bandno && \incground &  \rdground\\
      CL0054\_860 & $   0.559$ &    \memyes &    \bandno &    \bandno &   \bandyes &    \bandno &    \bandrf &    \bandno &    \bandno &   \bandyes &    \bandno && \incground &  \rdground\\
      CL0054\_918 & $   0.557$ &    \memyes &    \bandno &    \bandno &    \bandno &    \bandno &    \bandrf &    \bandno &    \bandno &    \bandno &    \bandno && \incground &  \rdground\\
      CL0054\_966 & $   0.559$ &    \memyes &    \bandno &    \bandno &    \bandno &    \bandno &    \bandrf &    \bandno &    \bandno &    \bandno &    \bandno && \incground &  \rdground\\
      MS1054\_C01 & $   0.828$ &    \memyes &    \bandno &    \bandno &    \bandno &   \bandyes &    \bandno &    \bandno &    \bandno &    \bandno &    \bandrf &&    \inchst &     \rdhst\\
     MS1054\_1403 & $   0.813$ &    \memyes &    \bandno &    \bandno &    \bandno &   \bandyes &    \bandno &    \bandno &    \bandno &    \bandno &    \bandrf &&    \inchst &     \rdhst\\
     MS1054\_2011 & $   0.841$ &    \memyes &    \bandno &    \bandno &    \bandno &   \bandyes &    \bandno &    \bandno &    \bandno &    \bandno &    \bandrf &&    \inchst &     \rdhst\\
\hline
\end{tabular}
\end{minipage}
\end{table*}

As the imaging used in this study is collected from a number of
sources with differing coverage, the bands in which photometry is
available varies for each galaxy.  This is summarized in Table
\ref{tab:phot_info}.  

In order to minimise the uncertainty in converting to rest-frame
$B$-band magnitude (see \S \ref{sec:phot}), we use as a basis the
magnitude measured in an observed band corresponding closest to the
redshifted $B$-band.  Due to differing redshifts and band
availabilities, this band varies, and is therefore
indicated for each galaxy in Table \ref{tab:phot_info}.

Usually the inclination and disk scalelength are measured from HST
images if available.  However, there are occasions, usually due to an
unreliable GIM2D fit, where measurements from a ground-based image
have been used instead.  To document this, Table \ref{tab:phot_info}
also gives the source of the inclination and disk scalelength
measurements.

Note that our measurements have been taken in such a way as to
minimise the effect of heterogeneous source images.  For example,
colours are based on aperture magnitudes measured on seeing-matched
images, while disk scalelengths, and usually inclinations, are
measured using a technique which takes into account the pixel scale
and seeing.



\begin{thebibliography}{}

\bibitem[\protect\citeauthoryear{Abadi, Moore, \& 
Bower}{1999}]{AMB99} Abadi M.~G., Moore B., Bower R.~G., 
1999, MNRAS, 308, 947 

\bibitem[\protect\citeauthoryear{{Arag{\' o}n-Salamanca}, {Ellis}, {Couch} \&
  {Carter}}{{Arag{\' o}n-Salamanca} et~al.}{1993}]{AECC93}
{Arag{\' o}n-Salamanca} A.,  {Ellis} R.~S.,  {Couch} W.~J.,    {Carter} D.,  1993,
  MNRAS, 262, 764

\bibitem[\protect\citeauthoryear{Balogh, Navarro, \& 
Morris}{2000}]{BNM00} Balogh M.~L., Navarro J.~F., Morris 
S.~L., 2000, ApJ, 540, 113

\bibitem[\protect\citeauthoryear{{Balogh} et~al.,}{{Balogh}
  et~al.}{2004}]{Betal04}
{Balogh} M.,  et~al., 2004, MNRAS, 348, 1355

\bibitem[\protect\citeauthoryear{Bamford, Arag{\' o}n-Salamanca \& 
Milvang-Jensen}{Bamford et~al.}{2005}]{BamfordField}
{Bamford} S.~P., Arag{\' o}n-Salamanca A., Milvang-Jensen B., 2005, 
submitted to MNRAS [Paper~2]

\bibitem[\protect\citeauthoryear{Bekki}{1998}]{B98} Bekki 
K., 1998, ApJ, 502, L133 

\bibitem[\protect\citeauthoryear{Bekki}{1999}]{B99} Bekki 
K., 1999, ApJ, 510, L15 

\bibitem[\protect\citeauthoryear{Bekki \& 
Couch}{2003}]{BC03} Bekki K., Couch W.~J., 2003, ApJ, 596, 
L13 

\bibitem[\protect\citeauthoryear{Bekki, Couch, \& 
Shioya}{2002}]{BCS02} Bekki K., Couch W.~J., Shioya Y., 2002, 
ApJ, 577, 651 

\bibitem[\protect\citeauthoryear{{Bertin} \& {Arnouts}}{{Bertin} \&
  {Arnouts}}{1996}]{sextractor}
{Bertin} E.,  {Arnouts} S.,  1996, A\&AS, 117, 393

\bibitem[\protect\citeauthoryear{Biviano et 
al.}{1990}]{BGMM90} Biviano A., Giuricin G., Mardirossian F., 
Mezzetti M., 1990, ApJS, 74, 325 

\bibitem[\protect\citeauthoryear{{B{\" o}hm} et~al.,}{{B{\" o}hm}
  et~al.}{2004}]{Bohmetal04}
{B{\" o}hm} A.,  et~al., 2004, A\&A, 420, 97

\bibitem[\protect\citeauthoryear{Burkert}{1995}]{B95} 
Burkert A., 1995, ApJ, 447, L25 

\bibitem[\protect\citeauthoryear{{Butcher} \& {Oemler}}{{Butcher} \&
  {Oemler}}{1978}]{BO78}
{Butcher} H.,  {Oemler} A.,  1978, ApJ, 219, 18

\bibitem[\protect\citeauthoryear{{Couch} \& {Sharples}}{{Couch} \&
  {Sharples}}{1987}]{CS87}
{Couch} W.~J.,  {Sharples} R.~M.,  1987, MNRAS, 229, 423

\bibitem[\protect\citeauthoryear{{Couch}, {Barger}, {Smail}, {Ellis} \&
  {Sharples}}{{Couch} et~al.}{1998}]{CBSES98}
{Couch} W.~J.,  {Barger} A.~J.,  {Smail} I.,  {Ellis} R.~S.,    {Sharples}
  R.~M.,  1998, ApJ, 497, 188

\bibitem[\protect\citeauthoryear{Dale et al.}{2001}]{Dale01} 
Dale D.~A., Giovanelli R., Haynes M.~P., Hardy E., Campusano L.~E., 2001, 
AJ, 121, 1886 

\bibitem[\protect\citeauthoryear{de Jong}{1996}]{dJ96} de 
Jong R.~S., 1996, A\&A, 313, 45

\bibitem[\protect\citeauthoryear{{De Lucia}, {Kauffmann}, {Springel}, {White},
  {Lanzoni}, {Stoehr}, {Tormen} \& {Yoshida}}{{De Lucia}
  et~al.}{2004}]{dLKSWLSTY04}
{De Lucia} G.,  {Kauffmann} G.,  {Springel} V.,  {White} S.~D.~M.,  {Lanzoni}
  B.,  {Stoehr} F.,  {Tormen} G.,    {Yoshida} N.,  2004, MNRAS, 348, 333

\bibitem[\protect\citeauthoryear{{Dressler},}{{Dressler}}
  {1980}]{D80}
{Dressler} A., 1980, ApJ, 236, 351

\bibitem[\protect\citeauthoryear{{Dressler} \& {Gunn}}{{Dressler} \&
  {Gunn}}{1982}]{DG82}
{Dressler} A.,  {Gunn} J.~E.,  1982, ApJ, 263, 533

\bibitem[\protect\citeauthoryear{{Dressler} \& {Gunn}}{{Dressler} \&
  {Gunn}}{1983}]{DG83}
{Dressler} A.,  {Gunn} J.~E.,  1983, ApJ, 270, 7

\bibitem[\protect\citeauthoryear{{Dressler} \& {Gunn}}{{Dressler} \&
  {Gunn}}{1992}]{DG92}
{Dressler} A.,  {Gunn} J.~E.,  1992, ApJS, 78, 1

\bibitem[\protect\citeauthoryear{{Dressler} et~al.,}{{Dressler}
  et~al.}{1997}]{DOCSEBBPS97}
{Dressler} A.,  et~al., 1997, ApJ, 490, 577

\bibitem[\protect\citeauthoryear{{Dressler}, {Smail}, {Poggianti}, {Butcher},
  {Couch}, {Ellis} \& {Oemler}}{{Dressler} et~al.}{1999}]{DSPBCEO99}
{Dressler} A.,  {Smail} I.,  {Poggianti} B.~M.,  {Butcher} H.,  {Couch} W.~J.,
  {Ellis} R.~S.,    {Oemler} A.~J.,  1999, ApJS, 122, 51

\bibitem[\protect\citeauthoryear{Gentile et 
al.}{2004}]{GSKVK04} Gentile G., Salucci P., Klein U., Vergani 
D., Kalberla P., 2004, MNRAS, 351, 903

\bibitem[\protect\citeauthoryear{{Gioia}, {Shaya}, {Le Fevre}, {Falco},
  {Luppino} \& {Hammer}}{{Gioia} et~al.}{1998}]{GSLFFLH98}
{Gioia} I.~M.,  {Shaya} E.~J.,  {Le Fevre} O.,  {Falco} E.~E.,  {Luppino}
  G.~A.,    {Hammer} F.,  1998, ApJ, 497, 573

\bibitem[\protect\citeauthoryear{{Girardi} \& {Mezzetti}}{{Girardi} \&
  {Mezzetti}}{2001}]{GM01}
{Girardi} M.,  {Mezzetti} M.,  2001, ApJ, 548, 79

\bibitem[\protect\citeauthoryear{Gnedin}{2003a}]{G03a} Gnedin 
O.~Y., 2003, ApJ, 582, 141 

\bibitem[\protect\citeauthoryear{Gnedin}{2003b}]{G03b} Gnedin 
O.~Y., 2003, ApJ, 589, 752 [G03b]

\bibitem[\protect\citeauthoryear{G{\' o}mez et 
al.}{2003}]{Gomez03} G{\' o}mez P.~L., et al., 2003, ApJ, 584, 
210 

\bibitem[\protect\citeauthoryear{{Goto} et~al.,}{{Goto}
  et~al.}{2003}]{Getal03}
{Goto} T.,  et~al., 2003, PASJ, 55, 757

\bibitem[\protect\citeauthoryear{{Gunn} \& {Gott}}{{Gunn} \&
  {Gott}}{1972}]{GG72}
{Gunn} J.~E.,  {Gott} J.~R.,  1972, ApJ, 176, 1

\bibitem[\protect\citeauthoryear{{Heavens}, {Panter}, {Jimenez} \&
  {Dunlop}}{{Heavens} et~al.}{2004}]{HPJD04}
{Heavens} A.,  {Panter} B.,  {Jimenez} R.,    {Dunlop} J.,  2004, Nature, 428,
  625

\bibitem[\protect\citeauthoryear{Henriksen \& 
Byrd}{1996}]{HB96} Henriksen M., Byrd G., 1996, ApJ, 459, 82 

\bibitem[\protect\citeauthoryear{{Hoekstra}, {Franx}, {Kuijken} \& {van
  Dokkum}}{{Hoekstra} et~al.}{2002}]{HFKvD02}
{Hoekstra} H.,  {Franx} M.,  {Kuijken} K.,    {van Dokkum} P.~G.,  2002,
  MNRAS, 333, 911

\bibitem[\protect\citeauthoryear{J{\" a}ger et 
al.}{2004}]{Jager04} J{\" a}ger K., Ziegler B.~L., B{\" o}hm 
A., Heidt J., M{\" o}llenhoff C., Hopp U., Mendez R.~H., Wagner S., 2004, 
A\&A, 422, 907

\bibitem[\protect\citeauthoryear{{Kashikawa} et~al.,}{{Kashikawa}
  et~al.}{2002}]{FOCAS}
{Kashikawa} N.,  et~al., 2002, PASJ, 54, 819

\bibitem[\protect\citeauthoryear{Lewis et al.}{2002}]{Lewis02} 
Lewis I., et al., 2002, MNRAS, 334, 673 

\bibitem[\protect\citeauthoryear{Metevier \& 
Koo}{2004}]{MK04} Metevier A.~J., Koo D.~C., 2004, IAUS, 
220, 415 

\bibitem[\protect\citeauthoryear{Metropolis et al.}{1953}]
{MRRTT53} Metropolis, N., Rosenbluth, A., Rosenbluth, M.,
Teller, A., Teller, E., 1953, J. Chem. Phys., 21, 1087 

\bibitem[\protect\citeauthoryear{Mihos \& 
Hernquist}{1994}]{MH94} Mihos J.~C., Hernquist L., 1994, 
ApJ, 425, L13 

\bibitem[\protect\citeauthoryear{{Milvang-Jensen}}{{Milvang-Jensen}}{2003}]{MJ%
thesis}
{Milvang-Jensen} B.,  2003, PhD thesis, University of Nottingham

\bibitem[\protect\citeauthoryear{{Milvang-Jensen}, {Arag{\' o}n-Salamanca},
  {Hau}, {J{\o}rgensen} \& {Hjorth}}{{Milvang-Jensen} et~al.}{2003}]{MJetal03}
{Milvang-Jensen} B.,  {Arag{\' o}n-Salamanca} A.,  {Hau} G.~K.~T.,
  {J{\o}rgensen} I.,    {Hjorth} J.,  2003, MNRAS, 339, L1

\bibitem[\protect\citeauthoryear{Moore et al.}{1998}]{M98} 
Moore B., Governato F., Quinn T., Stadel J., Lake G., 1998, ApJ, 499, L5 

\bibitem[\protect\citeauthoryear{Moore et al.}{1999}]{MLQS99} 
Moore B., Lake G., Quinn T., Stadel J., 1999, MNRAS, 304, 465 

\bibitem[\protect\citeauthoryear{Navarro \& 
Steinmetz}{2000}]{NS3} Navarro J.~F., Steinmetz M., 2000, 
ApJ, 538, 477 

\bibitem[\protect\citeauthoryear{Navarro, Frenk, \& 
White}{Navarro et~al.}{1996}]{NFW96} Navarro J.~F., Frenk C.~S., White 
S.~D.~M., 1996, ApJ, 462, 563

\bibitem[\protect\citeauthoryear{Owen et al.}{2005}]{OLKWM05} Owen, F.~N., Ledlow, 
M.~J., Keel, W.~C., Wang, Q.~D., Morrison, G.~E.\ 2005, AJ, 129, 31 

\bibitem[\protect\citeauthoryear{{Persic} \& {Salucci}}{{Persic} \&
  {Salucci}}{1991}]{PS91}
{Persic} M.,  {Salucci} P.,  1991, ApJ, 368, 60

\bibitem[\protect\citeauthoryear{{Pierce} \& {Tully}}{{Pierce} \&
  {Tully}}{1992}]{PT92}
{Pierce} M.~J.,  {Tully} R.~B.,  1992, ApJ, 387, 47 [PT92]

\bibitem[\protect\citeauthoryear{{Poggianti}, {Smail}, {Dressler}, {Couch},
  {Barger}, {Butcher}, {Ellis} \& {Oemler}}{{Poggianti}
  et~al.}{1999}]{PSDCBBEO99}
{Poggianti} B.~M.,  {Smail} I.,  {Dressler} A.,  {Couch} W.~J.,  {Barger}
  A.~J.,  {Butcher} H.,  {Ellis} R.~S.,    {Oemler} A.~J.,  1999, ApJ, 518,
  576

\bibitem[\protect\citeauthoryear{{Poggianti}, {Bridges}, {Komiyama}, {Yagi},
  {Carter}, {Mobasher}, {Okamura} \& {Kashikawa}}{{Poggianti}
  et~al.}{2004}]{PBKYCMOK04}
{Poggianti} B.~M.,  {Bridges} T.~J.,  {Komiyama} Y.,  {Yagi} M.,  {Carter} D.,
  {Mobasher} B.,  {Okamura} S.,    {Kashikawa} N.,  2004, ApJ, 601, 197

\bibitem[\protect\citeauthoryear{Quilis, Moore, \& 
Bower}{2000}]{QMB00} Quilis V., Moore B., Bower R., 2000, 
Sci, 288, 1617 

\bibitem[\protect\citeauthoryear{{Rudnick} et~al.,}{{Rudnick}
  et~al.}{2003}]{EDISCS_Messenger}
{Rudnick} G.,  et~al., 2003, The Messenger, 112, 19

\bibitem[\protect\citeauthoryear{Saha \& Williams}{1994}]{SW94}
Saha P., Williams T.~B., 1994, AJ, 107, 1295 

\bibitem[\protect\citeauthoryear{{Schlegel}, {Finkbeiner} \&
  {Davis}}{{Schlegel} et~al.}{1998}]{SFD98}
{Schlegel} D.~J.,  {Finkbeiner} D.~P.,    {Davis} M.,  1998, ApJ, 500, 525

\bibitem[\protect\citeauthoryear{{Seifert} et~al.,}{{Seifert}
  et~al.}{2000}]{FORS}
{Seifert} W., et~al., 2000, in Iye M., Moorwood, A.~F., eds, Proc. SPIE Vol. 4008, Optical and IR
  Telescope Instrumentation and Detectors, p. 96

\bibitem[\protect\citeauthoryear{{Simard} \& {Pritchet}}{{Simard} \&
  {Pritchet}}{1998}]{SP98}
{Simard} L.,  {Pritchet} C.~J.,  1998, ApJ, 505, 96

\bibitem[\protect\citeauthoryear{{Simard} \& {Pritchet}}{{Simard} \&
  {Pritchet}}{1999}]{SP99}
{Simard} L.,  {Pritchet} C.~J.,  1999, PASP, 111, 453

\bibitem[\protect\citeauthoryear{{Simard} et~al.,}{{Simard}
  et~al.}{2002}]{GIM2D}
{Simard} L.,  et~al., 2002, ApJS, 142, 1

\bibitem[\protect\citeauthoryear{{Smail}, {Dressler}, {Couch}, {Ellis},
  {Oemler}, {Butcher} \& {Sharples}}{{Smail} et~al.}{1997}]{SDCEOBS97}
{Smail} I.,  {Dressler} A.,  {Couch} W.~J.,  {Ellis} R.~S.,  {Oemler} A.~J.,
  {Butcher} H.,    {Sharples} R.~M.,  1997, ApJS, 110, 213

\bibitem[\protect\citeauthoryear{Spergel et 
al.}{2003}]{WMAP} Spergel D.~N., et al., 2003, ApJS, 148, 
175 

\bibitem[\protect\citeauthoryear{{Stocke}, {Morris}, {Gioia}, {Maccacaro},
  {Schild}, {Wolter}, {Fleming} \& {Henry}}{{Stocke} et~al.}{1991}]{SMGMSWFH91}
{Stocke} J.~T.,  {Morris} S.~L.,  {Gioia} I.~M.,  {Maccacaro} T.,  {Schild} R.,
   {Wolter} A.,  {Fleming} T.~A.,    {Henry} J.~P.,  1991, ApJS, 76, 813

\bibitem[\protect\citeauthoryear{Tran et al.}{2003}]{Tran03} 
Tran K.~H., Franx M., Illingworth G., Kelson D.~D., van Dokkum P., 2003, 
ApJ, 599, 865

\bibitem[\protect\citeauthoryear{{Tran}, {Franx}, {Illingworth}, {van Dokkum},
  {Kelson} \& {Magee}}{{Tran} et~al.}{2004}]{TFIvDKM04}
{Tran} K.~H.,  {Franx} M.,  {Illingworth} G.~D.,  {van Dokkum} P.,  {Kelson}
  D.~D.,    {Magee} D.,  2004, ApJ, 609, 683

\bibitem[\protect\citeauthoryear{{Tully} \& {Fouque}}{{Tully} \&
  {Fouque}}{1985}]{TF85}
{Tully} R.~B.,  {Fouque} P.,  1985, ApJS, 58, 67

\bibitem[\protect\citeauthoryear{van Albada et 
al.}{1985}]{vA85} van Albada T.~S., Bahcall J.~N., Begeman 
K., Sancisi R., 1985, ApJ, 295, 305 

\bibitem[\protect\citeauthoryear{{van Dokkum}, {Franx}, {Fabricant},
  {Illingworth} \& {Kelson}}{{van Dokkum} et~al.}{2000}]{vDFFIK00}
{van Dokkum} P.~G.,  {Franx} M.,  {Fabricant} D.,  {Illingworth} G.~D.,
  {Kelson} D.~D.,  2000, ApJ, 541, 95

\bibitem[\protect\citeauthoryear{{Vogt}, {Haynes}, {Giovanelli} \&
  {Herter}}{{Vogt} et~al.}{2004}]{Vogt04II}
{Vogt} N.~P.,  {Haynes} M.~P.,  {Giovanelli} R.,    {Herter} T.,  2004, AJ,
  127, 3300

\bibitem[\protect\citeauthoryear{{White} et~al.,}{{White}
  et~al.}{2005}]{EDISCS}
{White} S.~D.~M.,  et~al., 2005, submitted to A\&A

\bibitem[\protect\citeauthoryear{{Yang}, {Zabludoff}, {Zaritsky}, {Lauer} \&
  {Mihos}}{{Yang} et~al.}{2004}]{YZZLM04}
{Yang} Y.,  {Zabludoff} A.~I.,  {Zaritsky} D.,  {Lauer} T.~R.,    {Mihos}
  J.~C.,  2004, ApJ, 607, 258

\bibitem[\protect\citeauthoryear{{Ziegler}, {B{\" o}hm}, {J{\" a}ger}, {Heidt}
  \& {M{\" o}llenhoff}}{{Ziegler} et~al.}{2003}]{ZBJHM03}
{Ziegler} B.~L.,  {B{\" o}hm} A.,  {J{\" a}ger} K.,  {Heidt} J., 
{M{\" o}llenhoff} C., 2003, ApJ, 598, L87

\end{thebibliography}
\end{document}